\documentclass[prb,showpacs,preprintnumbers,preprint]{revtex4}
\usepackage{graphicx}
\begin{document}
\def\deltav{{\mbox{\boldmath{$\delta$}}}}
\def\rhov{{\mbox{\boldmath{$\rho$}}}}
\def\tauv{{\mbox{\boldmath{$\tau$}}}}
\def\Lambdav{{\mbox{\boldmath{$\Lambda$}}}}
\def\sigmav{{\mbox{\boldmath{$\sigma$}}}}
\def\xiv{{\mbox{\boldmath{$\xi$}}}}
\def\chiv{{\mbox{\boldmath{$\chi$}}}}
\def\rhov{{\mbox{\boldmath{$\rho$}}}}
\def\phiv{{\mbox{\boldmath{$\phi$}}}}
\def\piv{{\mbox{\boldmath{$\pi$}}}}
\def\psiv{{\mbox{\boldmath{$\psi$}}}}
\def\oh{{\scriptsize 1 \over \scriptsize 2}}
\def\ot{{\scriptsize 1 \over \scriptsize 3}}
\def\of{{\scriptsize 1 \over \scriptsize 4}}
\def\tf{{\scriptsize 3 \over \scriptsize 4}}
\title{Charge and Spin Ordering in the Mixed Valence Compound
LuFe$_2$O$_4$}
\author{A. B. Harris$^1$ and T. Yildirim$^2$}

\affiliation{[1] Department of Physics and Astronomy,
University of Pennsylvania, Philadelphia, PA 19104}
\affiliation{[2] NIST Center for Neutron Research, National Institute of
Standards and Technology, Gaithersburg, MD 20899 and
Department of Materials Science and Engineering,
University of Pennsylvania, Philadelphia, PA 19104}
%%% ----------------------------------------------------------------------
\date{\today}

\begin{abstract}
Landau theory and symmetry considerations lead us to propose an
explanation for several seemingly paradoxical behaviors
of charge ordering (CO) and spin ordering (SO) in the mixed valence compound
LuFe$_2$O$_4$.  Both SO and CO
are highly frustrated.  We analyze a lattice gas model of CO
within mean field theory and determine the magnitude of several of the
phenomenological interactions.  We show that the assumption of a
continuous phase transitions at which CO or SO
develops implies that both CO and SO are incommensurate.
To explain how ferroelectric fluctuations in the charge disordered phase
can be consistent with an {\it anti}ferroelectric ordered phase, we invoke
an electron-phonon interaction in which a low energy (20 meV) zone-center
transverse phonon plays a key role.  The energies of all the zone center
phonons are calculated from first principles.  We give a Landau analysis
which explains SO and we discuss a model of interactions
which stabilizes the SO state, if it assumed commensurate.
However, we suggest a high resolution experimental determination
to see whether this phase is really commensurate, as believed
up to now.  The applicability of representation analysis is discussed.
A tentative explanation for the sensitivity of the CO state to an
applied  magnetic field in field-cooled experiments is given.
\end{abstract}
\maketitle

\section{INTRODUCTION}

The phenomenon of charge ordering (CO) has been studied ever
since the observation of the Verwey transition[\onlinecite{VERWEY}]
in Fe$_3$O$_4$ (in which the average valence of the Fe ions is
$8/3$). An oft-cited paper by Anderson[\onlinecite{PWA}] proposed
a simple and appealing model to explain the ferroelectric charge
ordering.  However, the most recent high-precision neutron scattering
results[\onlinecite{JPW}] show that this model does not correctly
explain the CO in Fe$_3$O$_4$ and the explanation of the actual
nature of its CO remains an open question.[\onlinecite{KHOMSKII}]
Similarly $R$Fe$_2$O$_4$, where $R$ is a trivalent rare-earth and in
which the average valence of the Fe$^{+2}$ and Fe$^{3+}$ ions is $5/2$
presents an even more challenging problem to our understanding of CO.
In this paper we will consider LuFe$_2$O$_4$ (LFO) whose CO and magnetic
structure has been widely studied in recent
years.[\onlinecite{IIDA, IKEDA1, IKEDA2, YAMADA, EPS, ZHANG,
CHRIST, ANGST,WEN}].

\begin{figure}[h!]
\begin{center}
\vspace {0.4 in}
\includegraphics[width=5.0 cm]{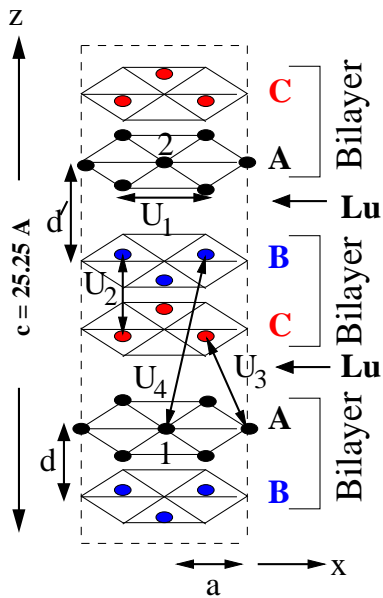}
\caption{\label{FIG1} (Color online)
Left: Fe ions in LFO.  The brackets indicate a bilayer consisting
of two Fe RLL's.  The presence of a TLL of Lu ions between adjacent
Fe bilayers is indicated.  The oxygen ions (not shown) are
almost uniformly distributed over the structure.  The hexagonal
(conventional) unit cell contains three bilayers configured so
that the TLL's are stacked in the order ABCABC (A=red, B=black, C=blue)
with two sites in the rhombohedral unit cell labeled "1" 
and "2" which are related by a center of inversion symmetry.  The
screened Coulomb interactions $U_n$ used below in our calculations
are indicated. $d=2.52\AA$ and $d'=5.81\AA$.}
\end{center}
\end{figure}

\begin{figure}[h!]
\begin{center}
\includegraphics[width=8.5 cm]{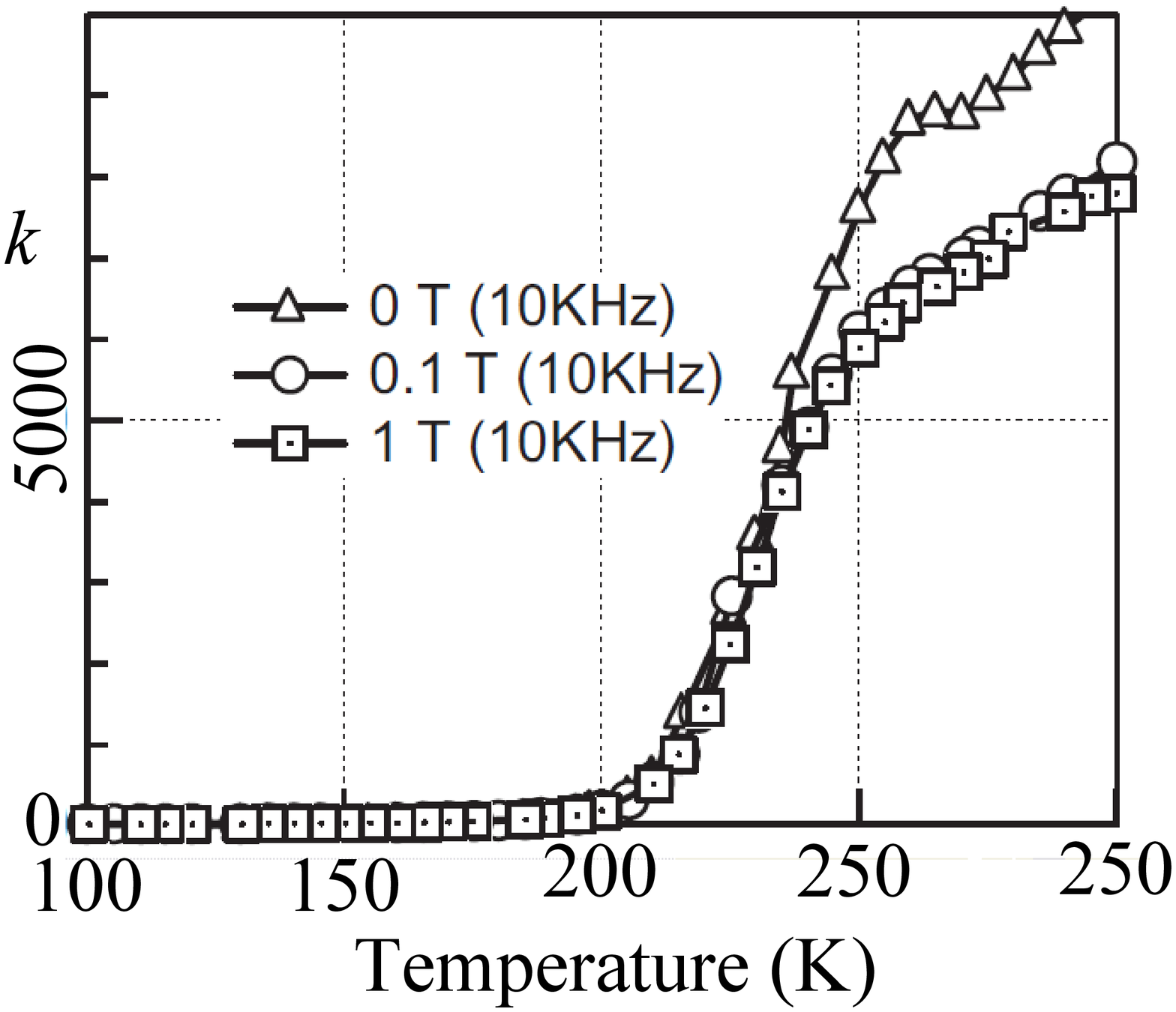}
\caption{\label{FIG2} The dielectric constant at zero frequency 
from Subramanian {\it et al}[\onlinecite{EPS}].}
\end{center}
\end{figure}

As an introduction we review the most salient experimental results
relevant to LFO.  In Fig. \ref{FIG1} we show the trigonal lattice structure
$R\overline 3 m$[\onlinecite{ACTA,ITC}] of LFO.  The generators of the
point group are a) inversion about the center of the unit cell, b) the
$x$-$y$ mirror plane, and c) the three-fold axis.  Note that the Fe ions form
triangular lattice layers (TLL's) arranged in bilayers.  The bilayers
are separated by a TLL of Lu ions.  The stacking of the Fe TLL's
is in the same order as for an fcc crystal.  The rhombohedral unit
cell spans three bilayers and there are two Fe ions
per rhombohedral primitive unit cell, as shown in Fig. 1.  At 
temperatures above 500K, the valence electrons can thermally hop so that
effectively all Fe sites appear to have charge 2.5e.[\onlinecite{TANAKA}]
Consistent with this electronic mobility, the dielectric constant
at zero frequency (shown in Fig. \ref{FIG2}) is very large for $T>200$
K [\onlinecite{EPS}].

\begin{figure}[h!]
\begin{center}
\vspace {0.4 in}
\includegraphics[width=7.0 cm]{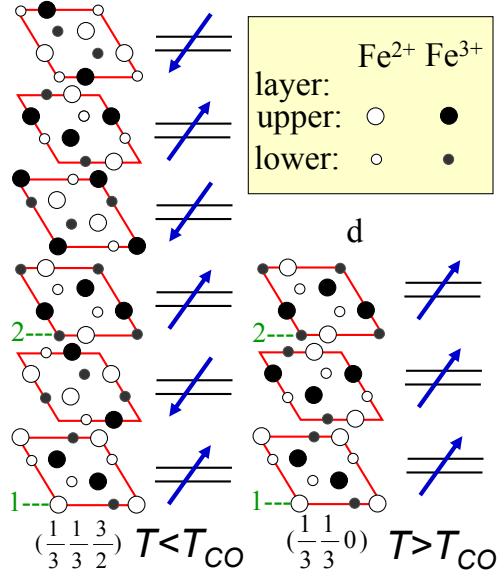}
\caption{\label{FIG2X} (Color online) From Ref. \onlinecite{ANGST}.  Fully
CO structures when the incommensurability of the wave vector is neglected.
The dipole moments of each bilayer (in the absence of incommensurability)
as calculated in Ref. \onlinecite{ANGST} are shown.  Left: The
`antiferroelectric' R3 structure.  Right the ``ferroelectric'  R3 structure.}
\end{center}
\end{figure}

As the temperature is reduced from 500K, CO correlations
develop at wave vectors which nearly coincide with ``root 3" (R3) ordering
(see Figs. 5 and 6, below)
within each TLL  and eventually at the charge ordering temperature
$T_{\rm CO} \approx 320$K three dimensional long range CO develops
via a continuous transition.[\onlinecite{WEN}]
Both the fluctuations and the long range order occur at incommensurate
values of the wave vector.[\onlinecite{YAMADA,ANGST}]
In the paraelectric phase ($T>T_{\rm CO}$)
the dominant fluctuations are consistent with no enlargement of the unit
cell in the ${\bf c}$ direction.[\onlinecite{ANGST}] We call such fluctuations
``ferro-incommensurate" (FI) fluctuations to emphasize that
their wave vector has
incommensurate in-plane components.  (The incommensurate in-plane components
are very close to the values of the X point of a two dimensional triangular
lattice gas with repulsive interactions, as we discuss below.)
Surprisingly, the CO that occurs for $T<T_{\rm CO}$ involves a doubling
of the unit cell along ${\bf c}$.[\onlinecite{ANGST,REY}]
We call this ordering ``antiferro-incommensurate" (AFI).
These structures are shown in Fig. \ref{FIG2X} where,
for simplicity, the small incommensurability of the wave vectors
is neglected.  A main objective of the present paper is to explain
why the CO phase is AFI and does not reflect the dominant FI 
fluctuations of the paraelectric (P) phase.

We will analyze this unusual CO within the lattice gas model
used by Yamada {\it et al}[\onlinecite{YAMADA}] which we refer to as Y.
The most striking result found by Y was that even with
in-plane coupling $U_1$ and interplane couplings $U_2$ and $U_3$,
long-range order is not possible because the maximum of the
wave vector dependent susceptibility occurs over an entire 
`degeneration line' in wave vector space. This result had been known
for similar spin and lattice gas models on a rhombohedral lattice
from the work of Rastelli and Tassi[\onlinecite{ENRICO1,ENRICO2}]
and later of Reimers and Dahn[\onlinecite{REIMERS}].  As Y found,
it was necessary to include an interaction ($U_4$ in Fig. 1) between
next-nearest neighboring TLL's in order to remove this degeneracy.
We will analyze this situation in detail and show that there
are two crucial parameters which govern this phenomenon.  The
first parameter is the interaction $U_2$ in Fig. \ref{FIG1}
which scales the radius of the cylinder on which the degeneration line
is wrapped. The second parameter
is the interaction $U_4$ in Fig. \ref{FIG1} which 
removes the continuous degeneracy and leads to an energy
variation of amplitude $\Delta E$ as one traverses the degeneration
line.  To elucidate our mean-field results we will briefly review the
results for a single TLL with repulsive interactions. Repulsive
interactions are clearly relevant for the charge-charge interactions.
For magnetic interactions the presence of many exchange paths through
intervening oxygen ions suggests that the magnetic interactions should
also be repulsive (i. e. antiferromagnetic).  The most general result
of our analysis is that the wave vector of the stacked TLL's of LFO
is unstable at X (the wave vector which characterizes the R3 structure)
if a continuous transition is assumed, in which case
the ordered phase must perforce be incommensurate.  

This same logic applies to the magnetic phase transition into the spin 
ordering (SO) phase which appears at $T=240$K.  We will discuss the
ramifications of the fact that the magnetic transition appears to be
a continuous one to a commensurately ordered spin state. We will give
a Landau analysis of the symmetry of the SO phase and will discuss
microscopic interactions which can explain this ordering.
Finally, we will discuss briefly a possible explanation of recent
field-cooled experiments[\onlinecite{WEN}] which show that such
a protocol has seems to significantly destabilize the AFI CO state.

\begin{figure}[h!]
\begin{center}
\vspace {0.4 in}
\includegraphics[width=9.0 cm]{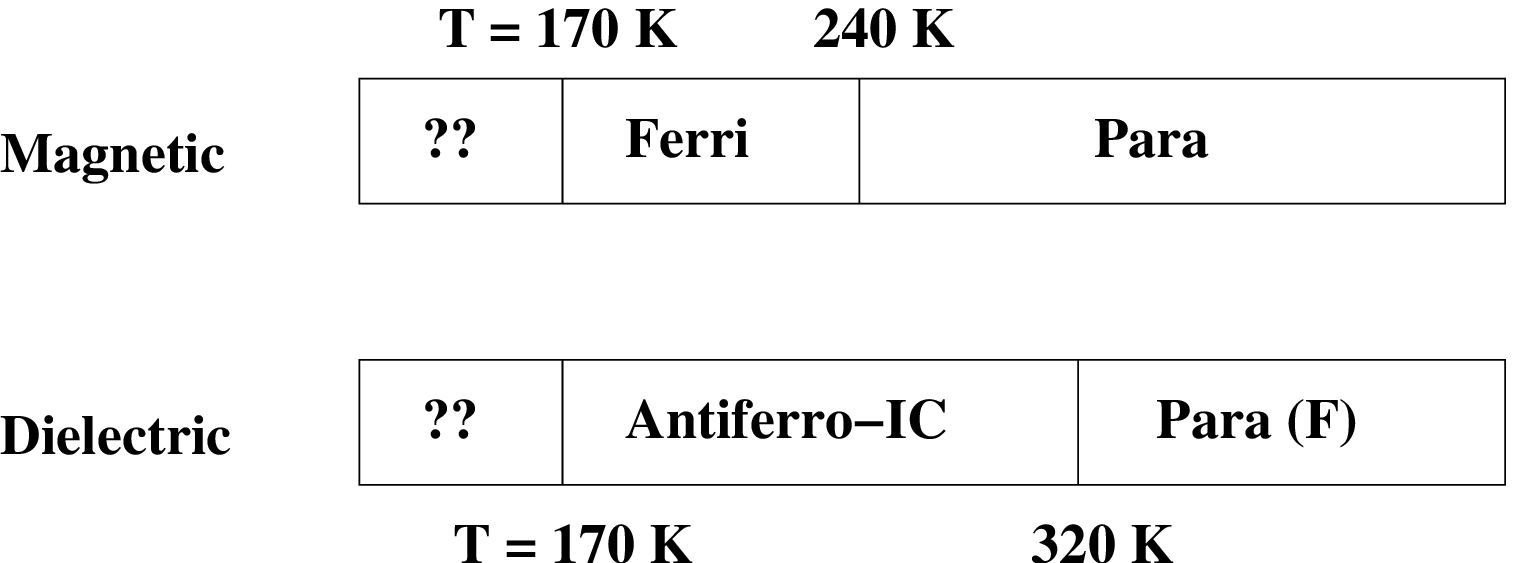}
\caption{\label{FIG3} The magnetoelectric phase diagram of LFO
based on the results discussed in the text.}
\end{center}
\end{figure}

Below 170K the system undergoes another transition in which the
magnetic order parameter sharply decreases.[\onlinecite{CHRIST}]
The details of this state are not settled at present[\onlinecite{WEN}]
and we will not consider it further.  Thus the magnetoelectric 
phase diagram we are considering is that shown in Fig.  \ref{FIG3}.
A brief summary of this work appeared some time ago.[\onlinecite{PRE}]

\section{MEAN FIELD TREATMENT OF CO}

\subsection{Calculation}

We start with a Landau analysis of CO using the lattice gas model of
Y in which one introduces a variable $\sigma_n({\bf R})$, where $n$ ($n=1,2$)
labels the $n$th site of the rhombohedral unit cell at ${\bf R}$.
which assumes the value $+1$ ($-1$) if the site is occupied
by an Fe$^{3+}$ (Fe$^{+2}$) ion. Then $x_n({\bf R})\equiv \langle
\sigma_n({\bf R})\rangle$, where $\langle \ \ \rangle$ is a thermal average.
As shown in Fig. 1, we include an
interaction $U_1$ between nearest neighbors within the same TLL,
an interaction $U_2$ between nearest neighbors in adjacent TLL's
within the same bilayer, an interaction $U_3$ between nearest
neighbors in adjacent bilayers, and an interaction $U_4$ between
nearest neighbors in second-neighboring TLL's.  As argued by Y,
in view of the large dielectric constant (See Fig. \ref{FIG2})
we prefer to use a model in which the interactions fall off rapidly
with separation rather than one based on a long-ranged Coulomb
interaction[\onlinecite{NAGPRL}].  The free energy is written
in terms of the Fourier transformed variables
\begin{eqnarray}
x_n({\bf q}) &=& \sum_{\bf R} x_n ({\bf R}) \exp (i {\bf q}
\cdot {\bf R}) \ .
\end{eqnarray}
Then
\begin{eqnarray}
F &=& \frac{1}{2} \sum_{\bf q} \sum_{n,m=1}^2
F_{nm}({\bf q}) x_n({\bf q})^* x_m({\bf q})
+ {\cal O} ( x^4 ) \ ,
\label{EQA} \end{eqnarray}
where $F_{nm}({\bf q})=F_{mn}({\bf q})^*$.  The free energy must be
invariant under spatial inversion ${\cal I}$, since ${\cal I}$ is a
symmetry of this lattice.  Under spatial inversion ${\cal I}$, site 
\#1 goes into site \#2, so that ${\cal I}S_1({\bf q}) = S_2({\bf q})^*$,
from which we conclude that $F_{11}=F_{22}$.  A continuous CO  
transition is signaled by the appearance of a zero eigenvalue
of the quadratic form.  This instability will first occur at a
wave vector whose value we wish to determine.

\begin{figure}[h!]
\begin{center}
\vspace {0.4 in}
\includegraphics[width=14.0 cm]{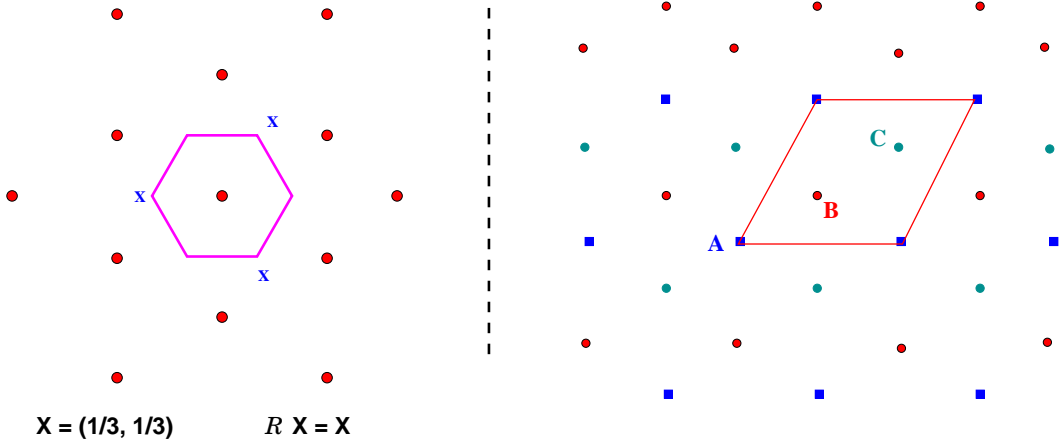}
\caption{\label{FIG4L} (Color online) Left: The reciprocal lattice
of the TLL.  The X points all equivalent to one another under a
three-fold rotation ${\bf R}$, as discussed in the text. Right:  the
R3 structure associated with the wave vector of the X point.
The amplitudes $Z$ of the three sites within the R3 unit cell are given by
$Z_A = Z \cos(\phi)$, $Z_B = Z \cos (\phi+ 2 \pi /3)$, and
$Z_C = Z \cos( \phi+ 4 \pi /3)$.  The choice $\phi=0$ yields
$(Z_A,Z_B,Z_C) \propto (2,-1,-1)$ and the choice $\phi=\pi/6$
yields $(Z_A,Z_B,Z_C) \propto (1,0,-1)$.}
\end{center}
\end{figure}

\begin{figure}[h!]
\begin{center}
\includegraphics[width=8.0 cm]{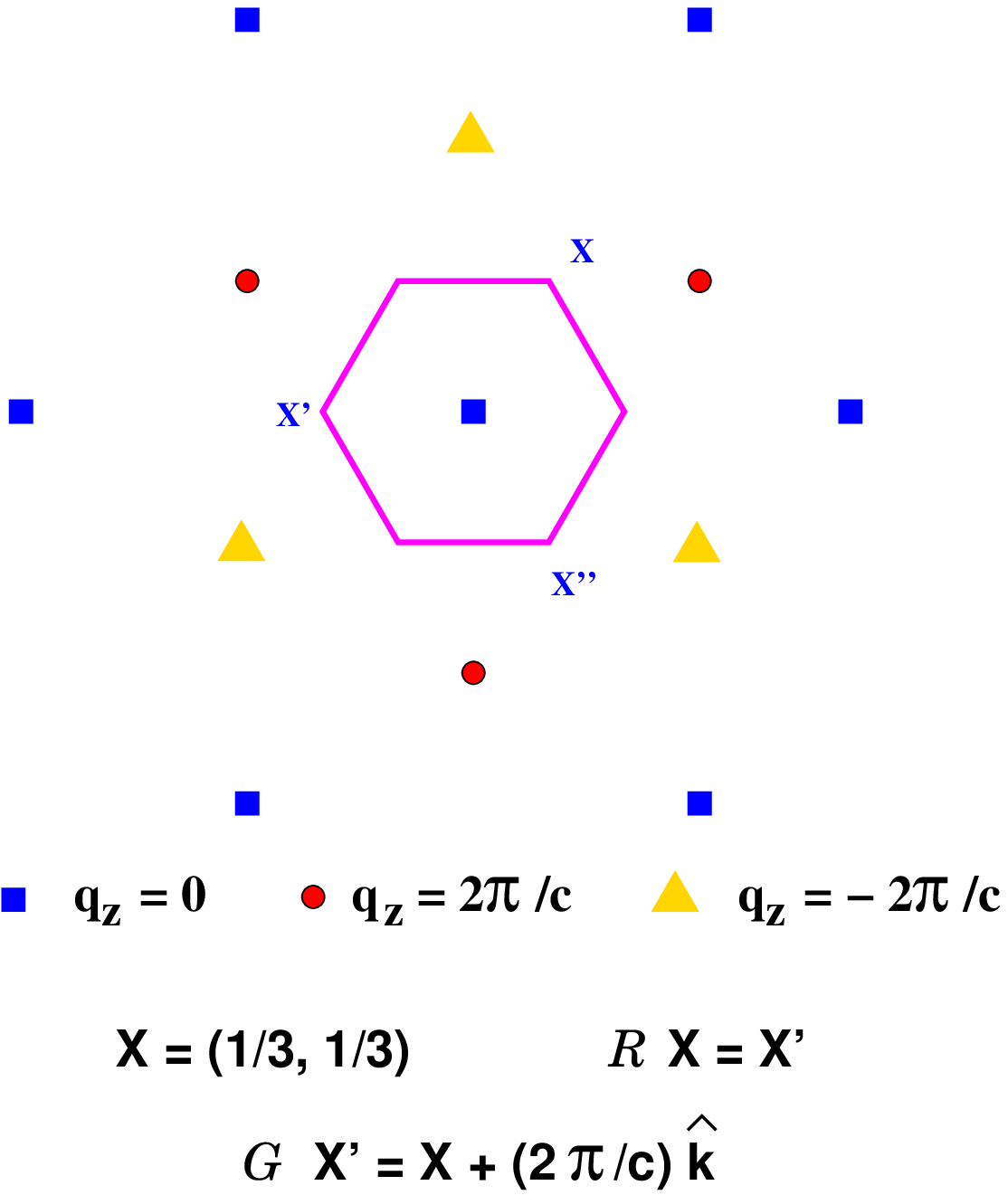}
\caption{\label{FIG4R} (Color online) The reciprocal lattice of the
rhombohedral lattice.}
\end{center}
\end{figure}

Before analyzing the model in detail we review some results for the simpler
problem of a lattice gas with repulsive interactions on a single TLL. In
left panel of Fig. \ref{FIG4L} we show the first Brillouin zone
for the TLL with the X points which are the wave vectors of the
ordered phase of this system. Note that the X point (which gives
rise to the CO or SO structure, shown in the right panel of Fig.
\ref{FIG4L}), is an isolated point having higher symmetry than that of 
all surrounding points.  To see this, note that uniquely this  point 
is invariant under a three-fold rotation about the center of zone,
because under the three-fold rotation such a point is taken into
another X point which is equal (modulo a reciprocal lattice vector)
to to the original point.  As a consequence of the special symmetry
of the $X$ point, its wave vector,
determined by the instability in the quadratic term of the
free energy, is {\it stable} with respect to the addition of small
further neighbor interactions.[\onlinecite{STOKES}]
Such structures have been observed
many times over the last half century.[\onlinecite{GRAFOIL}] In 
contrast, consider the analogous X points in the rhombohedral 
reciprocal lattice shown in Fig. \ref{FIG4R}. 
(We label these points as X regardless of the value of $q_z$
and refer to them as an ``X line.")  
Here, an X point is not invariant under a threefold
rotation because the points before and after such a rotation are
not equal modulo a reciprocal lattice vector. (The point is that
the rhombohedral reciprocal lattice vector does not connect points
before and after a three-fold rotation because the reciprocal lattice
vector needed to relate the components in the plane does not have
$q_z=0$.)   So if a transition at this wave vector is continuous,
the wave vector must perforce be incommensurate. Thus, without any
calculation, we have shown that Landau theory explains why the CO
wave vector is incommensurate. (How this conclusion relates to
representation theory is discussed in Appendix A.) Accordingly, the
CO phase should {\it not} display a spontaneous polarization, ${\bf P}$.
The nonzero value of ${\bf P}$ may be an artifact of the small electric
field applied during the experiment (as argued in Ref. \onlinecite{ANGST}).
Also, it is possible that the pyroelectric current (from which the
value of ${\bf P}$ is deduced) might be confused with currents
which develop in the conductive sample.[\onlinecite{WEN2}]
At a recent conference it was reported that ${\bf P}=0$ at low
temperature.  We will discuss below that a similar analysis of the
magnetic phase transition at $T=240$K indicates that if that transition
is continuous the ordered phase ought to be incommensurate.

We now return to the explicit calculation of the incommensurate CO
wave vector.  As mentioned in the introduction, from previous work
[\onlinecite{ENRICO1,ENRICO2,REIMERS,YAMADA}] it is known that
the minimal model that gives stable three-dimensional long-range order
requires the interactions shown in Fig. 1.  Other interactions
(such as a second-neighbor in-plane interaction or a second neighbor
interaction between adjacent TLL's) only lead to perturbative corrections.
Therefore we simplify the analysis by only considering the minimal model.
To determine the wave vector of CO within this model, we analyze the
appearance of a zero eigenvalue of the quadratic form of Eq. (\ref{EQA}).
In mean field theory one writes
\begin{eqnarray}
F_{nm}({\bf q}) = c'kT \delta_{nm} + \sum_{\bf R} U(0,m;{\bf R},n)
\exp(i {\bf q} \cdot {\bf R}) \ ,
\label{EQB} \end{eqnarray}
where $c'$ is a constant of order unity and $U(0,m;{\bf R},n)$ is
the interaction between sites $m$ in the rhombohedral unit cell
at the origin and $n$ in the rhombohedral unit cell at ${\bf R}$.  
We set $c'=1$, and $k_B=1$, and henceforth, unless stated otherwise,
all energies will be in temperature units. In Cartesian coordinates
\begin{eqnarray}
F_{11}&=& T + U_1 [ 2\cos(aq_x) + 4 \cos(aq_x/2) \cos(\sqrt 3 aq_y/2)] 
\nonumber \\ &&  + U_4 \left[ e^{icq_z/3} \Lambda(q_x,q_y)
+ e^{-icq_z/3} \Lambda(q_x,q_y)^* \right]
\nonumber \\ 
F_{21}&=& U_2 e^{-2icq_z/3} \Lambda(q_x,q_y)
% \nonumber \\ && \hspace{0.4 in}
+ U_3 e^{-icq_z/3} \Lambda(q_x,q_y)^* \ ,
\label{EQF12} \end{eqnarray}
where 
\begin{eqnarray}
\Lambda(q_x,q_y) &=& 2 e^{iq_y a \sqrt 3 /6} \cos(aq_x/2) +
e^{-iaq_y \sqrt 3 /3} \ .
\end{eqnarray}
To organize the calculation we will consider $U_n/U_1$ for $n>1$
to be of order the expansion parameter $\lambda$ and we will
work consistently to the lowest sensible order in $\lambda$,
keeping in mind that for $U_4=0$ we have a line of infinite
degeneration[\onlinecite{YAMADA,ENRICO1,ENRICO2,REIMERS}].
In any event these works indicate that for
$U_1 >0$ the instability in the quadratic form first appears near
the $X$ point, for some discrete values of $q_z$, for $U_4 \not= 0$
and for all $q_z$ for $U_4=0$.  Accordingly, we set
$aq_x = 4 \pi /3 + \rho_x$ and $aq_y=\rho_y$ and determine
$\rho$ to leading order in $\lambda$ for arbitrary $q_z$.

For this purpose we evaluate the matrix $F_{nm}$.  We find that
up to order $U_1\lambda^2$,
\begin{eqnarray}
F_{11} &=& T + U_1[ -3 + (3/4) \rho^2 ] - U_4 \sqrt 3 \rho \cos(\phi+ck_z/3)
\ ,
\end{eqnarray}
where $\rho \cos \phi = \rho_x$ and $\rho \sin \phi = \rho_y$,
with $\rho >0$, and $\rho^2=\rho_x^2+\rho_y^2$.  As we shall see
below, $\rho = {\cal O}(\lambda)$ and to clarify the situation it is
only necessary to calculate the eigenvalues of the quadratic form
${\bf F}$ to order $\lambda^2U_1$.
\begin{eqnarray}
\Lambda = - (\sqrt 3 /2) \rho e^{i \phi} \ .
\end{eqnarray}

Thus the critical eigenvalue, $\mu({\bf q})$  which first
approaches zero as the transition is approached is given by
$F_{11} - |F_{12}|$ or, up to order $U_1\lambda^2$,
\begin{eqnarray}
\mu({\bf q}) &=&
T + U_1[ -3 + (3/4) \rho^2 ] - U_4 \sqrt 3 \rho \cos(\phi+cq_z/3)
\nonumber \\ && - \frac{\sqrt 3 \rho }{2}
\Bigl[ U_2^2 + U_3^2 + 2 U_2U_3 \cos (2 \phi - cq_z/3) \Bigr]^{1/2} \ .
\label{EQ6} \end{eqnarray}

Note that for a single TLL, for which $U_2=U_3=U_4=0$ the X point with $\rho=0$
is stable and that, in view of the term linear in $\rho$, the X point becomes
unstable in the presence of interlayer interactions.[\onlinecite{LAMBDA}]
One might have expected to have three dimensional long-range order when
$U_1$, $U_2$, and $U_3$ are all nonzero because then each TLL's
interacts with ones above and below it.  However, the special symmetry
of the rhombohedral lattice prevents ordering[\onlinecite{ENRICO1,ENRICO2,
REIMERS,YAMADA}] when only these interactions are present, so we are
forced to include a nonzero value of $U_4$.

We first minimize $\mu({\bf q})$ with respect to $\rho$, which
is determined by
\begin{eqnarray}
0 &=& \frac{\partial \mu}{\partial \rho} = - \sqrt 3 U_4 \cos(\phi+cq_z/3)
+ \frac{3}{2} U_1 \rho - \frac{\sqrt 3}{2} [
U_2^2 + U_3^2 + 2U_2U_3 \cos(2 \phi- cq_z/3)] \ .
\end{eqnarray}
Thus  the value of $\rho$ which minimizes $\mu$ and which we denote
$\rho^*$ is given by
\begin{eqnarray}
\rho^* &=& (2 \sqrt 3 /3)(U_4/U_1) \cos(\phi+cq_z/3) \nonumber \\ &&
+ \frac{\sqrt 3}{3U_1} [ U_2^2 + U_3^2 + 2U_2U_3 \cos(2 \phi- cq_z/3)] \ .
\label{RHOEQ} \end{eqnarray}
As we mentioned, $\rho^*$ becomes nonzero at order $\lambda$.
Since $\partial \mu / \partial \rho =0$ at the extremum, corrections
to $\rho^*$ at the next order in $\lambda$ do not affect the result we
find for the critical eigenvalue $\mu(q_z, \phi, \rho^*(q_z,\phi))$:
\begin{eqnarray}
\mu (q_z,\phi) &=& T - 3U_1 -\sqrt 3 \rho^* U_4 \cos(\phi+cq_z/3)
+ (3/4) U_1 (\rho^*)^2 \nonumber \\ &&
- (\sqrt 3 \rho^*/2)[U_2^2 + U_3^2 + 2U_2U_3 \cos(2\phi-cq_z/3)]^{1/2} 
\nonumber \\ &=& T -3U_1- (U_4^2/U_1)\cos^2 (\phi_+)
-(U_4X/U_1) \cos(\phi_+) - X^2/(4U_1) \ ,
\end{eqnarray}
where $\phi_+=\phi+ck_z/3$ and
\begin{eqnarray}
X &=& \left[ U_2^2 + U_3^2 + 2U_2U_3 \cos(2\phi-cq_z/3) \right]^{1/2} \ .
\end{eqnarray}
When $U_4=0$, the critical eigenvalue is a function of the variable
$2\phi-cq_z/3$ and the eigenvalue is minimal for
$2 \phi-cq_z/3=0$ or $2 \pi$ if $U_2U_3>0$ and for 
$2 \phi-cq_z/3=\pi$ or $-\pi$ if $U_2U_3<0$, consistent with the results
cited for the line of degeneration.

\begin{figure}[h!]
\begin{center}
\includegraphics[width=5.0 cm]{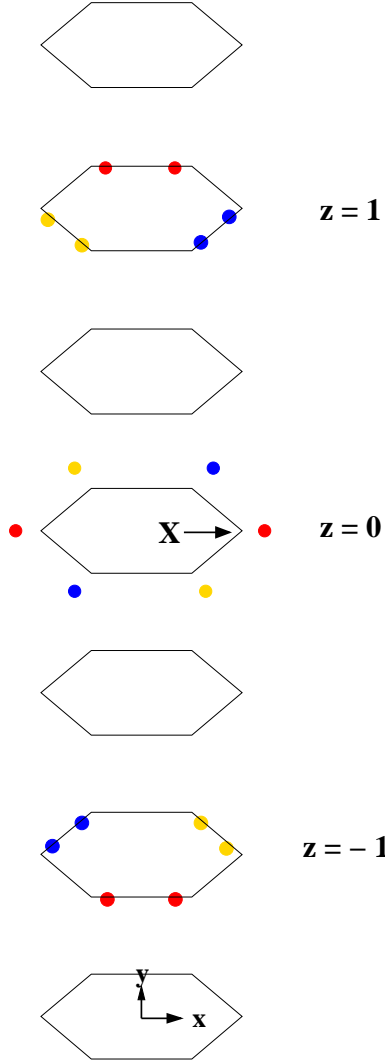}
\caption{\label{SCAT1} (Color online)
Scattering for positive $U_2U_3$ and positive $U_4$.  The $x$ and $y$ axes
of the scattering vector $\vec q$ are indicated. Its $z$-component is
$q_z=2 \pi z/c$.  The $X$ point, which in hexagonal notation is
$(1/3,1/3,0)$, is indicated.  Here we show the diffuse scattering in the
charge disordered phase (where the colors have no significance.) If
the system were to order, one would have Bragg scattering from
each of the three domains.  The scattering from a single domain
is indicated by a single color (red, blue, or gold). Note
that the diffuse scattering exhibits all the symmetries of the
crystal, whereas a single domain has lower symmetry, since it does
not have three fold symmetry.  However, if all domains are equally
populated, then the three fold symmetry is restored. This is the
result for the crystal shown in Fig. 1. Note that the mirror which 
takes $y$ into $-y$ is {\it not} a symmetry of the crystal and it
leads to a slightly different diffraction pattern.}
\end{center}
\end{figure}

\begin{figure}[h!]
\begin{center}
\includegraphics[width=5.0 cm]{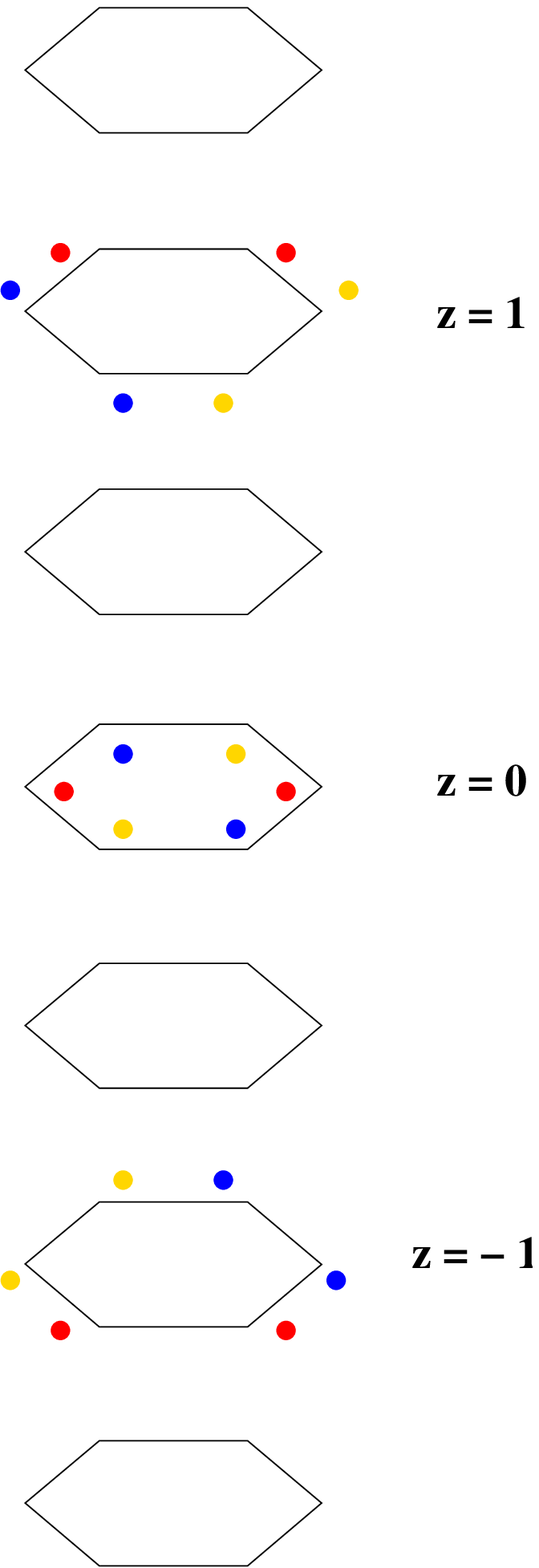}
\caption{\label{SCAT2} (Color online)
As Fig. \ref{SCAT1}, but for positive $U_2U_3$ and negative $U_4$.}  
\end{center}
\end{figure}

\begin{figure}[h!]
\begin{center}
\includegraphics[width=5.0 cm]{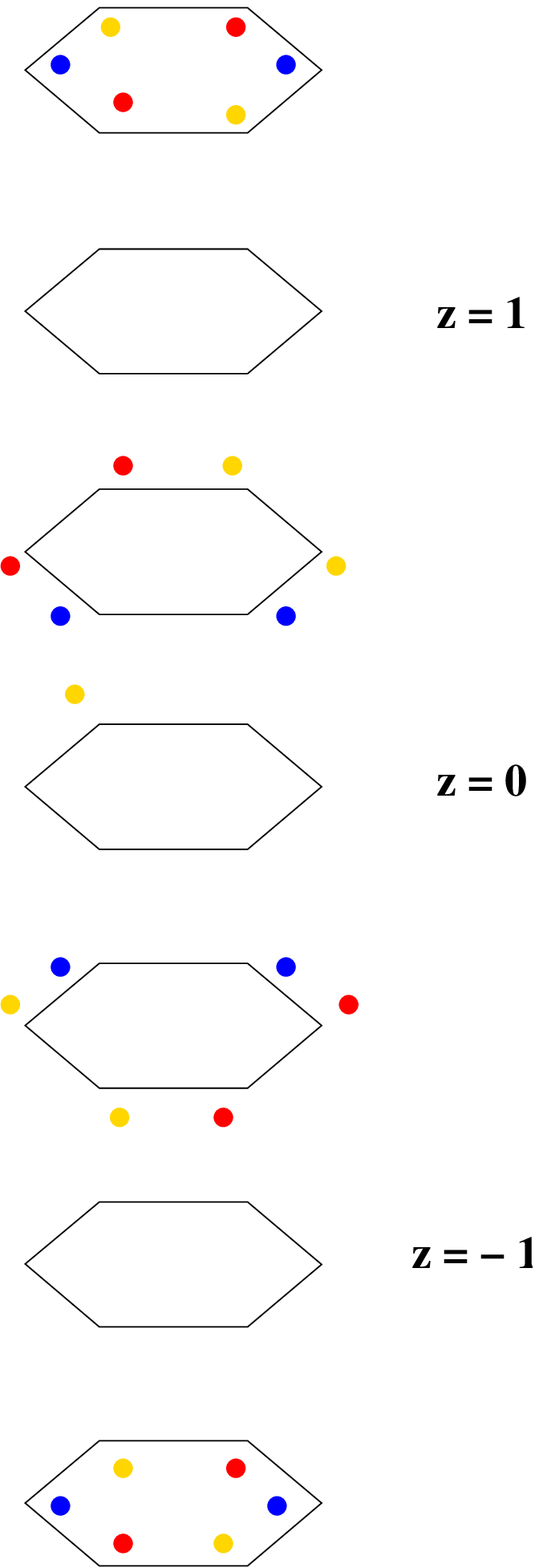}
\caption{\label{SCAT3} (Color online)
As Fig. \ref{SCAT1}, but for negative $U_2U_3$ and positive $U_4$.}  
\end{center}
\end{figure}

\begin{figure}[h!]
\begin{center}
\includegraphics[width=5.0 cm]{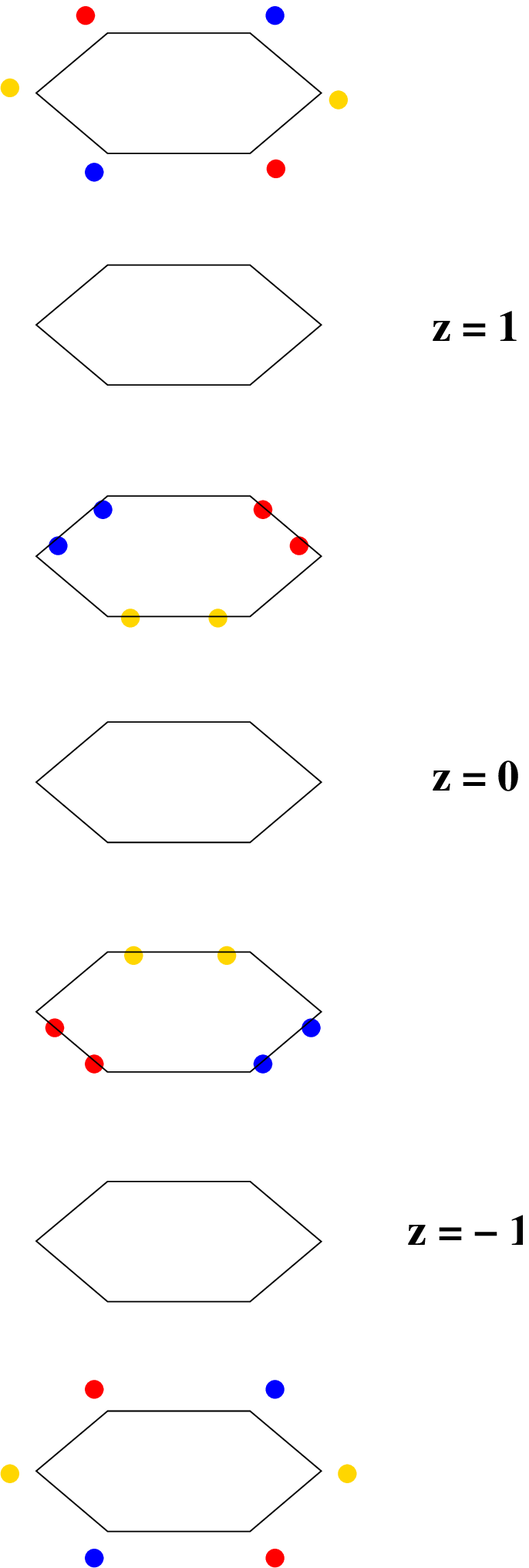}
\caption{\label{SCAT4} (Color online)
As Fig. \ref{SCAT1}, but for negative $U_2U_3$ and negative $U_4$.}  
\end{center}
\end{figure}

The values of $\phi$ and $cq_z$ which complete the determination of the
critical wave vector when $U_4\not= 0$ are selected as those which minimize
$\mu$.  If we define
\begin{eqnarray}
n &\equiv& U_4/|U_4| \ , \hspace {1 in}
m \equiv U_2U_3/|U_2U_3| \ ,
\end{eqnarray}
then we see that $\mu$ is minimized by setting
\begin{eqnarray}
\cos(\phi+cq_z/3)=n\ , \hspace{1 in} \cos(2 \phi-cq_z/3) = m \ ,
\end{eqnarray}
so that
\begin{eqnarray}
\phi + cq_z/3 &=& (1-n) \pi /2 + 2 k \pi \ , \hspace{1 in}
2 \phi - cq_z/3 = (1-m) \pi /2 + 2l \pi \ ,
\end{eqnarray}
where the integers $k$ and $l$ are free parameters. We therefore get
the results of Table \ref{phik}, shown in Figs. \ref{SCAT1}-\ref{SCAT4}. 

\begin{table}
\caption{\label{phik} The critical wave vector as a function of
the $U$'s is given by $\vec q = (\rho \cos \phi, \rho \sin phi, q_z)$,
where $\rho$ is given by Eq. (\ref{RHOEQ}) and $\phi$ and $q_z$ are
given below, where $p$ is an integer.} 
\vspace{0.2 in}
\begin{tabular} {||c| c| c| c ||} \hline
\ \ $\frac{U_2U_3}{|U_2U_3|}$ \ \ & \  \ $\frac{U_4}{|U_4|}$
\ \ & \ \ $\phi$\ \ & \ \ $cq_z$\ \ \\
\hline
+ & + & $\frac{2p\pi}{3}$ & $- 2 p \pi$ \\
+ & - & $\frac{(2p+1)\pi}{3}$ & $2 \pi (-p+1)$ \\
- & + & $\frac{(2p+1)\pi}{3}$ &  $2 \pi (-p-1/2)$\\
- & - & \ \ $\frac{2\pi(p+1)}{3}$\ \  &\ \  $2 \pi(-p+1/2)$\ \ \\
\hline \end{tabular}
\end{table}

\subsection{Comparison to Experiment}

\begin{figure}[h!]
\begin{center}
\vspace{0.4 in}
\includegraphics[width=5.0 cm]{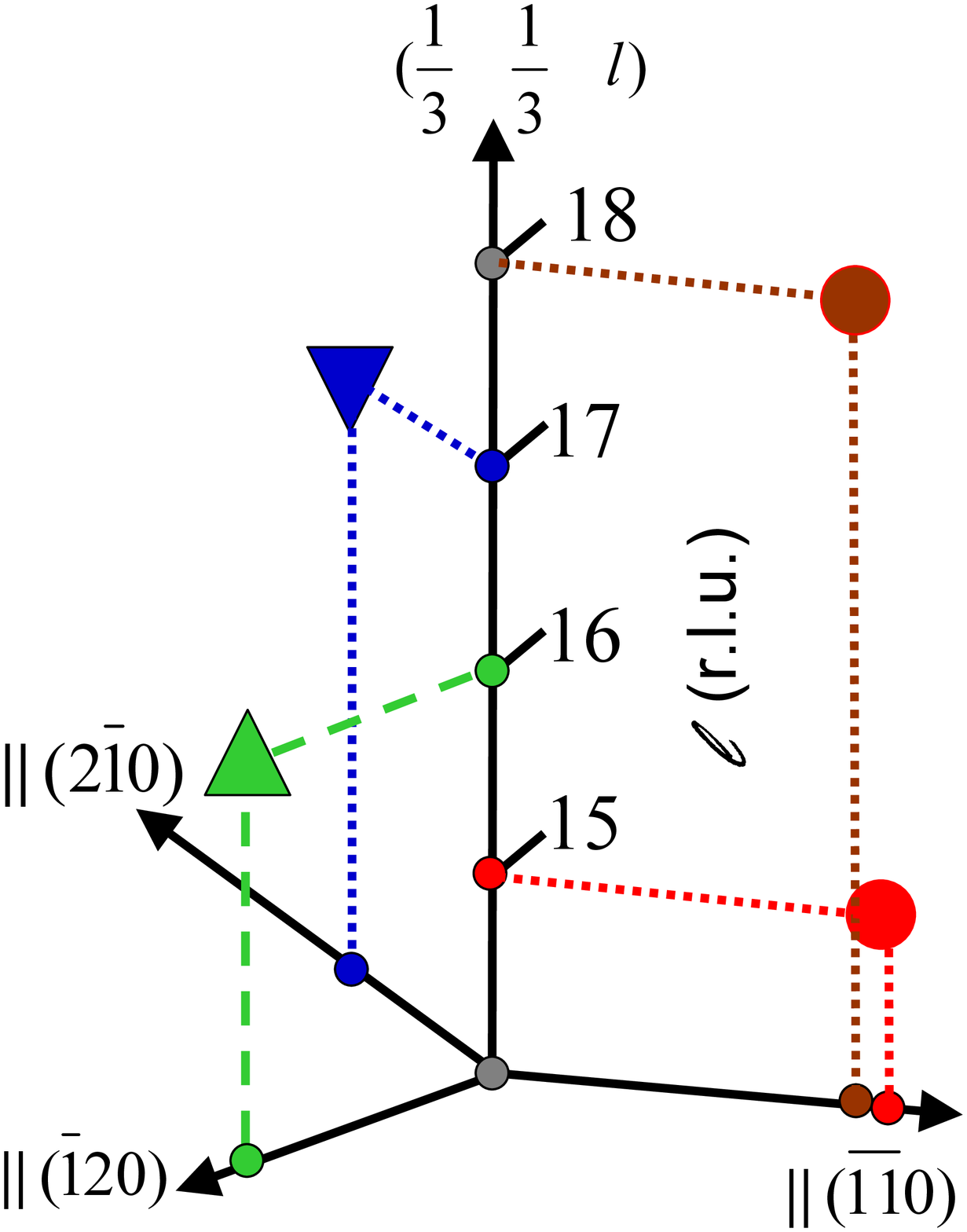}
\caption{\label{ADATA} (Color online)
From Ref. \onlinecite{ANGST}.  Diffraction maxima in the charge disordered
phase ($T=360$K) near an X line.  Hexagonal coordinates are used. Different
colors correspond to wave vectors of different domains when CO occurs.}
\end{center}
\end{figure}

Now we compare these results with experiment and if we obtain
agreement we should be able to identify some of the parameters.
Look at the data shown in Fig. \ref{ADATA}.  Note that the diffraction
pattern at $l=15$ (or $l=18$) can be compared with that
for $l=0$ in Figs. \ref{SCAT1} to \ref{SCAT4}. Note that the data indicate
that the diffraction at $l=15$ occurs for $\phi=\pi$ [since it is closer
to $(0,0,l)$ than is the commensurate location]. So this is  the same as
shown in Fig. \ref{SCAT2} for $U_2U_3>0$ and $U_4<0$ and can also be confirmed
by comparison with that case (for $p=1$) in Table I. Furthermore, as
one moves in the direction of positive $l=q_z$, one sees that
$d\phi/dl= - 2\pi /3$
in both the experiment and in Fig. \ref{SCAT2}. (Although this line shows a
definite sign of helicity, the system as a whole is not chiral.  Of the
six lines equivalent to the one shown in Fig. \ref{ADATA}, three of them have
one sign of helicity and three the other sign of helicity.) We can also check
that our eigenvector of the matrix $\bf F$ of Eq. (\ref{EQA}) agrees with 
that used in Ref. \onlinecite{ANGST}. They use $[1,-1]$ for FI ($l=0$)
diffraction (see the first paragraph of p3).  For us to obtain that result
$F_{12}$ must be real positive (to give a minimal eigenvalue).  This implies
that $\phi=\pi$ which agrees with Table I.  
It is more problematic to connect the AFI diffraction to our analysis
because the AFI phase can not be explained by the present theory, although
the AFI diffraction is that shown in Fig. \ref{SCAT4}.

Now we fix the parameters to fit the existing data. In this connection
we will assume that $U_1 \gg U_2 \gg U_3 \gg U_4$,  as will be justified
a posteriori.  Then the mean-field value of the CO transition
temperature, which we denote $T_{\rm CO,MF}$
is determined by setting $\mu({\bf q})=0$.  Since we
believe that the heavily screened interactions decay
rapidly with distance, this condition gives
\begin{eqnarray}
T_{\rm CO,MF}= 3U_1 \ .
\end{eqnarray}
If we were to identify this with the observed CO transition at
$T_{\rm CO}= 320$K, then we would conclude that $U_1=110$ K.
But since the coupling between bilayers
is very weak, the two dimensional fluctuations will cause the observed
value of $T_{\rm CO}$ to be very much less than $T_{\rm CO,MF}$. 
Accordingly we adopt the estimate[\onlinecite{rho2}]
\begin{eqnarray}
U_1 &=& 500 {\rm K} \ .
\end{eqnarray}
Next, we have to decide whether to take $U_2U_3$ to
be positive, as suggested by the fact that the P phase diffraction
dominantly occurs at integer values of $cq_z/(2\pi)$[\onlinecite{ANGST}]
or to be negative, as suggested by the fact that the CO phase diffraction
occurs at half-integer values of 
$cq_z/(2 \pi)$.[\onlinecite{ANGST,IKEDA2,YAMADA}]
Since the corrections to our mean field theory are smallest in the
P phase, we use the P phase data to fix $U_2U_3>0$ and hope
to explain the CO data by some correction to this theory.
(Of course, in addition, if we took $U_2U_3<0$, we would have
to explain why screening causes $U_2$ or $U_3$ to be negative.)
Equation (\ref{RHOEQ}) gives[\onlinecite{rho}] 
\begin{eqnarray}
U_2 = \sqrt 3 \rho U_1 = 0.06 U_1 = 30 {\rm K} \ .
\end{eqnarray}
In view of the large ratio $U_1/U_2=15$, it seems reasonable to get
$U_2$ under the assumption that $U_3/U_2 \ll 1$.  We do not have
an unambiguous way to determine $U_3$ and $U_4$.  However, considering
that $U_2/U_1$ is about 15, we guess that $U_2/U_4=15$, which
would indicate that $|U_4|=2$K. Previously we determined that to fit
the diffraction at $T=360$K we needed to assume that $U_4$ was negative,
so we set
\begin{eqnarray}
U_4 = - 2 {\rm K} \ .
\end{eqnarray}

\subsection{FI versus AFI Transition Temperatures}

Now we want to estimate the difference between the mean-field values of
the transition temperatures for FI and AFI CO
which we denote respectively as $T_{\rm MF,F}$ and $T_{\rm MF,AF}$.
(By our choice of parameters $\Delta T \equiv T_{\rm MF,F}-T_{\rm MF,AF}$
is positive, which does not agree with the experimental value.
We have that
\begin{eqnarray}
\Delta T &=& \mu(cq_z/\pi) - \mu(cq_z=0) \ ,
\end{eqnarray}
when these $\mu$'s have each been minimized with respect to $\phi$
and $\rho$. Accordingly we need
\begin{eqnarray}
0 &=& \frac{\partial \mu(\phi, q_z)}{\partial \phi} = \frac{2U_4^2}{U_1}
\cos(\phi_+)\sin(\phi_+) + \frac{U_4X}{U_1} \sin(\phi_+)
\nonumber \\ && 
+ \frac{U_2U_3}{2XU_1} [ X + 2 U_4\cos(\phi_+)] \sin(2\phi-cq_z/3)
\label{MINEQ} \end{eqnarray}
For $cq_z=0$, this minimization gives $\phi=\pi$, so that
\begin{eqnarray}
\mu(cq_z=0) &=& T - 3U_1 - U_4^2/U_1 + U_4(U_2+U_3)/U_1
- (U_2+U_3)^2/(4U_1) \ .
\end{eqnarray}

Now we analyze the extremum of $\mu$ for $cq_z= \pi$.
All terms in Eq. (\ref{MINEQ}) are of order $\lambda^2U_1$.  So we
assume that $U_4/U_2$ and $U_4/U_3$ are small and work to first
order in those quantities.  We then find that Eq. (\ref{MINEQ}) yields
that the extremum occurs for $\phi=\phi^*$, where
\begin{eqnarray}
\phi^* &=& \frac{\pi}{6} - \frac{U_4}{U_3} \equiv \frac{\pi}{6} 
+ \delta \phi \ .
\end{eqnarray}
Then
\begin{eqnarray}
\mu (cq_z=\pi) &=& T - 3U_1 - \frac{(U_2+U_3)^2}{4U_1} + {\cal O} (U_4^2)
\end{eqnarray}
and
\begin{eqnarray}
\Delta T &=& \frac{U_4^2}{U_1} 
- \frac{U_4}{U_1} (U_2+U_3) \rightarrow - \frac{U_2U_4}{U_1} \ .
\end{eqnarray}
Note that the implied negative sign for $U_4$ is crucial to explain
the dominance of FI fluctuations for $T>T_{\rm CO}$.
Using our admittedly arbitrary estimate of $U_4$ we have
\begin{eqnarray}
\Delta T &=& - \frac{U_4U_2}{U_1} = - \frac{(-2)(30)}{500} = 0.12 {\rm K} \ .
\end{eqnarray}
In view of the effect of large two dimensional fluctuations we estimate
that more realistically this model would give
\begin{eqnarray}
\Delta T &=& 0.04K \ .
\label{REAL} \end{eqnarray}
Of course, experiment[\onlinecite{ANGST}] tells us that $\Delta T < 0$
({\it i. e.} the first criticality we encounter as the temperature is
lowered is that toward the AFI phase) and below we will explain how
this can happen, even though $U_2U_3>0$.

\subsection{Summary}

Note that we used the amplitude $\rho$ of the incommensurate wave vector
to fix $U_2$, in contrast to the work of Y, who somehow uses this data
to fix $U_4$.  As we have said, the effect of $U_4$ is to scale the
amplitude of variation of the free energy as one traverses what, when
$U_4$ is zero, would be the degeneration line. In other words, $U_4$
scales $\Delta T$, the difference in the critical temperatures for
FI and AFI fluctuations and the negative sign of $U_4$ is crucial.
We find that $\Delta T$ is extremely small
because it is scaled by the long range interaction $U_4$ between
second neighboring TLL's.

\section{COMPETITION BETWEEN FI AND AFI STATES}

\begin{figure}[h!]
\begin{center}
\includegraphics[width=8.6 cm]{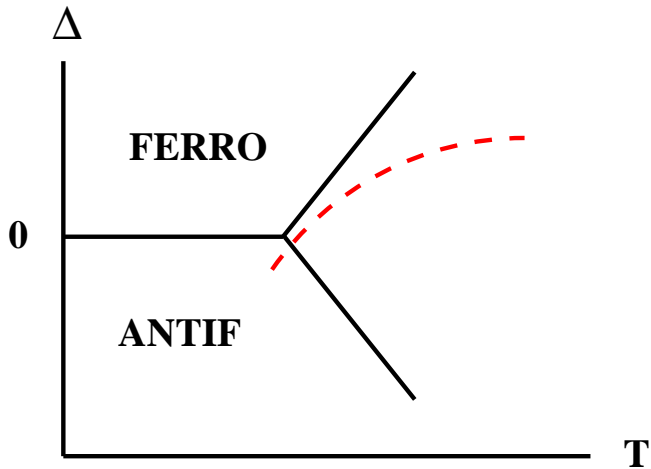}
\caption{\label{FIG6} (Color online) Mean-field phase 
diagram[\onlinecite{BRUCE}] near the bicritical point of
Eq. (\ref{EQBB}) for $v>0$. Since the dashed trajectory in the
disordered (P) phase is closer to the FI phase than to the
AFI phase, FI fluctuations dominate AFI fluctuations
in the P phase.}
\end{center}
\end{figure}

We now analyze the competition between FI ordering (at $q_z=0)$
and AFI ordering (at $q_z=3\pi /c$). Although the mean field
value of the transition temperature depends only very weakly on
$q_z$, it is simplest to invoke a model in which only
FI fluctuations at $\tilde q_z=0$ and AFI fluctuations at
$\tilde q_z =1/2$ compete.  Therefore we are led to consider the
model free energy of the form[\onlinecite{BRUCE}]
\begin{eqnarray}
F_0 &=& \frac{1}{2} (T-T_0 + \Delta /2)x_A^2 + \frac{1}{2} 
(T - T_0 - \Delta /2) x_F^2
+ u [x_A^2 + x_F^2]^2 + v x_A^2 x_F^2 \ ,
\label{EQBB} \end{eqnarray}
where $x_A$ ($x_F$) is the AFI (FI) order parameter.
The mean-field  temperature for AFI (FI) ordering is
$T_0 - \Delta /2$ ($T_0 + \Delta /2$) and $\Delta$ is positive
for $U_4<0$.  Then if, as is usually the case, $\Delta$ is
temperature independent, one would predict that as the temperature
is reduced, one would first enter the FI phase, which is not what we want.
So we propose a mechanism such that $\Delta$ is temperature-dependent,
so that as the temperature is decreased, we follow the dashed
trajectory on the phase diagram for this model[\onlinecite{BRUCE}]
shown in Fig. \ref{FIG6}. There is no reason to expect that
within the models considered thus far that
$\Delta$ (which arose from the value of $U_4$) should have a
relatively strong temperature dependence.[\onlinecite{EXPL}]
It is known[\onlinecite{RG}]
that the terms of order $x^4$ in Eq. (\ref{EQBB}) implement the
fixed length  constraint on the variables and lead to
a temperature dependent renormalization of the coefficients of
the quadratic terms.  But there is no reason to think that such
a renormalization will affect the AFI order parameter much more than
the FI order parameter. It has been suggested [\onlinecite{XU}]
that this anomalous crossover from FI to AFI fluctuations could
be explained by ``order from disorder."[\onlinecite{JV}] Here this
mechanism (of Ref. \onlinecite{NAGPRL}) relies on orbital fluctuations.
However, it would seem that the spin-orbit interaction would cause
the orbital degrees of freedom to be locked to whatever ordering occurs
in the spin degrees of freedom.  So we do not consider this mechanism.
While it is true that at zero temperature
quantum fluctuations exist in an antiferromagnet[\onlinecite{SHENDER}]
but are zero for a ferromagnet, we are too far from that regime
to invoke quantum fluctuations.  Similar effects do arise from thermal
fluctuations.[\onlinecite{JV}]  But here
the antiferromagnetic spin-wave energy is linear whereas the
ferromagnetic spin-wave energy is quadratic in wave vector.  Therefore
for identical coupling constants, ferromagnetic fluctuations have
lower energy than their  antiferromagnetic  counterparts.  This
argument suggests that FI fluctuations should be stronger than
AFI fluctuations.  Therefore we reject the suggestion[\onlinecite{XU}]
that the cross over from FI to AFI fluctuations can be attributed to
this mechanism, known as ``order from disorder."

Instead, to obtain the proposed trajectory shown in Fig. \ref{FIG6} we
invoke the coupling of the FI and AFI variables to a noncritical variable,
$Y$, so that the free energy is now $F=F_0+V$, where[\onlinecite{FN3}]
\begin{eqnarray}
V &=& a |x^2| Y + (1/2) \chi_Y^{-1} Y^2 \ ,
\label{VEQ} \end{eqnarray}
where here $x$ denotes either the FI or AFI order parameter.
Also $a$ is a temperature independent coupling constant and
$\chi_Y$ is the stiffness associated with $Y$ and
is almost temperature dependent because $Y$ is far from criticality.
(As we shall see, a suitable choice for $Y$ is a zone-center phonon.)
Since $Y$ is a noncritical variable we can eliminate it by
minimizing $F$ with respect to it, in which case we obtain
\begin{eqnarray}
F &=& F_0 - (1/2) \chi_Y a^2 |x^2|^2\ .
\label{FERROEQ} \end{eqnarray}
To leading order in the fluctuations we replace 
$|x^2|^2 \equiv {x({\bf q})^*}^2 x({\bf q})^2$ by
$4 x({\bf q}) x({\bf q})^* \langle x({\bf q})
x({\bf q})^* \rangle$, where $\langle Z \rangle$ is the
thermal expectation value of $Z$.[\onlinecite{RG}]
Then, if the coupling constant
for $FI$ fluctuations is $a_{FI}$ and that for AFI fluctuations is
$a_{AFI}$, this mechanism leads to the result
\begin{eqnarray}
\Delta(T) = \Delta - [a_{AFI}^2-a_{FI}^2] \chi_Y
\langle x({\bf q}) x({\bf q})^* \rangle \ ,
\label{EQFF} \end{eqnarray}
where we assume the thermal average is the same for FI and AFI
fluctuations.  So, if $a_{FI} \ll a_{AFI}$, then this mechanism 
leads to a renormalization of the quadratic term which is stronger 
for the AFI fluctuations than for the FI fluctuations. Then
the natural temperature dependence of the thermal average
of $|\sigma^2|$ can give the trajectory we desire.

\begin{figure}[h!]
\begin{center}
\vspace{0.4 in}
\includegraphics[width=12.0 cm]{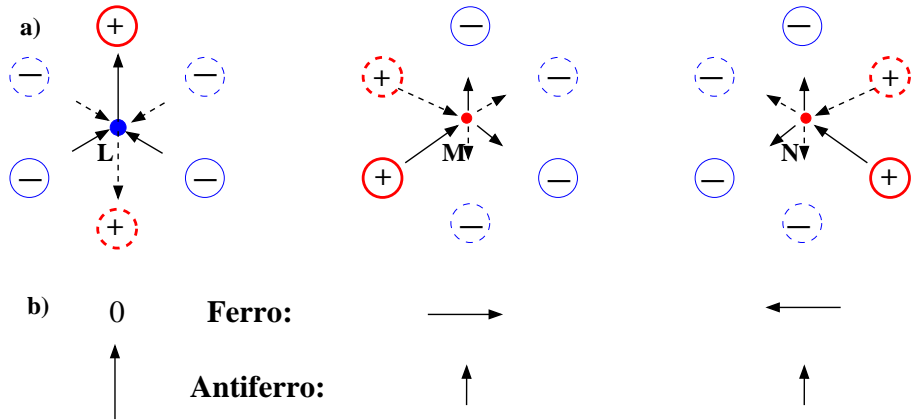}
\caption{(Color online) \label{FIG7} a) The in-plane components of
forces on sites L, M, and N in TLL$_0$ (as in Fig. 1) in the F configuration.
The solid line vector forces are from the charges (solid circles) in the
TLL above TLL$_0$ and  the dashed line vector forces are from the charges
(dashed circles) in the TLL below TLL$_0$.  The charges at L, M, and N
are negative, positive, and positive, respectively. (Red = positive,
blue = negative.) The larger
dots and heavier lined circles are charges of twice the magnitude
of the smaller dots and lighter lined circles. 
b) The net force on the sites assuming the separations between all
planes are the same.   For the AF configuration
the dashed forces are reversed and the resulting total forces are listed
as ``Antiferro:".  The forces are nonzero at zero wave
vector only for AF ordering.}
\end{center}
\end{figure}

If we choose $Y$ to be a zone-center phonon, the interaction we
consider is written schematically as
\begin{eqnarray}
V &=& \frac{1}{2} \sum_i M \omega_D u_i^2 + \sum_{ij}
u_{ji} \cdot \Bigl[ \nabla_r U(r_{ij}) Q_i Q_j \Bigr] \ ,
\label{EQCC} \end{eqnarray}
where $M \omega_D$ defines the Debye model, $u$ is a phonon displacement,
$U(r_{ij})$ is the heavily screened interaction, and $Q_i$ is the
charge on site $i$. The factor in Eq. (\ref{EQCC}) in square brackets 
is the force on site $i$ due to the charge on site $j$.

Since we will need the phonon energies, we implemented a 
a first-principles calculation of the energies of the zone-center
phonons in LuFe$_2$O$_4$. The calculations were performed within the plane-wave
implementation of the local density approximation to density functional
theory as implemented in the PWscf package.[\onlinecite{Baroni:1}]
We used Vanderbilt-type
ultrasoft potentials with Perdew-Zunger exchange correlation. A cutoff
energy of 408 eV and a 9$\times $9$\times $9 k-point mesh were found to be
enough for the total energy to converge within 0.5meV/atom.
The gamma phonon energies  were calculated using the supercell
method with finite difference.[\onlinecite{TY}] The primitive cell was
used and the full dynamical matrix was obtained from a total of eight
symmetry-independent atomic displacements ($\pm$0.02 {\AA}).

\begin{figure}
%\vspace{3.2 in}
\includegraphics[width=8.6cm]{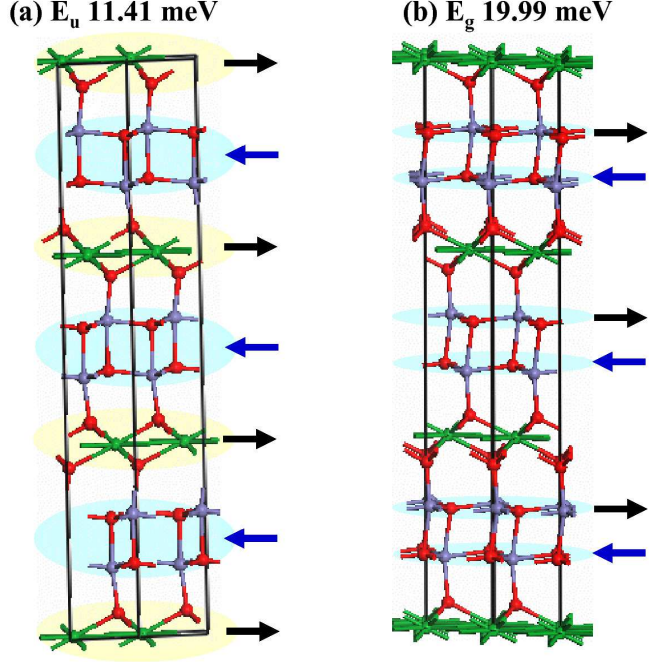}
\caption{ (Color online)
The displacements of the two lowest energy modes as discussed in the text.
Here O ions are red, Fe ions are purple, and  Lu ions are green.
The light blue ovals indicate units which move approximately rigidly.}
\label{fig15}
\end{figure}

The primitive cell contains one formula unit of  LuFe$_2$O$_4$,
giving rise to a total of 21 phonon branches. The phonon modes at 
$\Gamma $ are classified as
$\Gamma $ (q=0) = 4$A_{2u}$(IR) + 3$A_{1g}$(R) + 4$E_{u}$(IR) + 3$E_{g}$(R),
where R and IR correspond to Raman- and infrared-active, respectively. 
The nondegenerate (A) and doubly degenerate (E) modes correspond to motion
along the c-axis and within the ab-plane, respectively. In the Raman-active
modes the atoms at $(0,0,z)$ and $(0,0,-z)$ move out-of-phase (i.e. opposite)
while in the IR-modes they move in phase. 

The calculated mode energies and their symmetry labels are 
listed in Table \ref{PHON}.  We hope that our calculations
will initiate more experimental work such as Raman/IR measurements to
confirm the $\Gamma$ phonon energies that we calculated here.
From this table we see that the characteristic phonon energies are of
the order of 50 or more meV. Since the coupling to the phonons with
the lowest energies will be the most effective, we show the two lowest
energy phonons schematically in Fig. \ref{fig15}.  

The lowest energy mode with  E$_u$ symmetry and 11.4 meV energy corresponds
to displacements in the ab plane in which the LuO and FeO-bilayer move in
opposite directions as rigid units (see Fig. \ref{fig15}a). 
Hence the energy of this mode is basically determined by the strength
of LuO-Fe-O bond angle.  Even though this mode has the lowest energy,
its symmetry is not right to create the electrostatic 
forces needed for our mechanism. The next mode has the E$_g$ symmetry
and it corresponds to displacements in the ab-plane (Fig. \ref{fig15}b)
in which the two TTL's of each bilayer moves in opposite directions
while LuO-layer is fixed. 
Hence this modes involves twice as much O-Fe-O bond bending as the lowest
energy mode and interestingly it has about the twice energy (20 meV) of
the E$_u$ mode shown in Fig. \ref{fig15}a.  As we shall see, it is this
mode that creates the electrostatic forces needed for our mechanism.

\begin{table}[htbp]
\begin{center}
\caption{\label{PHON} List of phonon symmetries and calculated energies
(in meV) of LuFe$_2$O$_4$  at the $\Gamma $ point of the primitive cell, as 
obtained from the first-principles calculations described in the text.}
\vspace{0.2 in}
\begin{tabular}{|l|llllll|}
\hline\hline
Mode Symmety & E$_u$ & E$_g$ & A$_{2u}$ & A$_{1g}$ & A$_{2u}$ & E$_{u}$  \\
Energy (meV) &  11.41 & 19.99 & 20.02 & 31.48 & 38.46& 41.20 \\ \hline
Mode Symmety & A$_{1g}$ & E$_g$ & A$_{2u}$ & E$_{u}$ & E$_{g}$ & A$_{1g}$  \\
Energy (meV) & 53.25  & 54.73 & 57.69 & 58.79 & 62.62 & 84.96 \\ 
\hline\hline
\end{tabular}
\label{tab1}
\end{center}
\end{table}

Accordingly, we look for charge-phonon interactions which involve
zone-center transverse (to the $c$-axis) phonons.  We now analyze
the force on the Fe charges in a given TLL, which we denote TLL$_0$,
from the nearest neighboring Fe TLL's above and below TLL$_0$.
Since we consider coupling
to the lowest energy phonons, which involve motion transverse to
the ${\bf c}$ axis, we will only  consider forces in the
plane of the TLL. Phonon modes which decrease the distance between
TLL's will involve higher energy.  One sees that a low energy mode
which can couple the way we want is a zone-center phonon in which
alternate TLL's are displaced transversely relative to one another.
For this rhombohedral lattice, such a mode is an E$_g$ mode.
As mentioned this is the E$_g$ mode at 19.99 meV.  To see whether
this mode couples differently to FI and AFI fluctuations, we have only
to analyze the transverse force on one TLL from the TLL's above and
below it.  Since we are near the transition, we assume the R3
structure of the fluctuations (see Fig. \ref{FIG4L}, where we choose
$\phi=0$) and add up the forces in Fig. \ref{FIG7}. Also, we simplify
the argument by neglecting the fact that the interlayer separation is 
different for TLL's within the same bilayer and for TLL's in adjacent 
bilayers.

We now estimate quantitatively the effect of this coupling in Eq.
(\ref{EQCC}).  As in Fig. 1, the $Q$'s are given in terms
of the order parameter $x_{\rm X}$, where $X$
indicates either the FI or AFI configuration of TLL's.
Because the transverse motion of planes is relatively 
soft, we consider displacements to lie within the TLL$_0$ plane. 
When minimized with respect to $u_i$, the free energy is
\begin{eqnarray}
F_{\rm Ph} &=& - \frac{z_X^2 |x_X|^4} {2 M \omega_D^2}
\left( \frac{U}{r} \right)^2 \left( \frac{r_\parallel}{r} \right)^2
\xi^2 \ ,
\end{eqnarray}
where $\xi = (r/U) (dU/dr)$, $r_\parallel$ is the component of
$r$ within the TLL, and $z$ is the effective number of nearest
neighbors.  Also we set $U = (U_2+U_3)/2 = 16.5$K $\approx$ 1.5 meV.
The actual number of nearest neighbors is 6, but since
the forces do not all add up, we take $z=3$ for the AF configuration
and $z=0$ for the F configuration where the forces from
adjacent TLL's nearly cancel.  We set $\hbar \omega_D = 20$ meV,
$r=5\AA$, and $r_\parallel=2\AA$.  The factor $\xi$ depends
on how rapidly the interaction decreases with distance.  For bare
Coulomb interactions $\xi=-1$.  But we are far from that regime.  We
take $\xi=-10$, which is a value often found for exchange interactions
in insulators.[\onlinecite{BLOCH}] Also the Fe mass is $M=60$ amu, so its
reduced mass is 30 amu.  Thus
\begin{eqnarray}
F_{\rm Ph} &=& - F_0 |x_{\rm AF}^4|  \rightarrow
- \frac{1}{2} \Bigl( 8 F_0 \langle | x_{\rm AF}|^2 \rangle
\Bigr) | x_{\rm AF}|^2 \ ,
\end{eqnarray}
where $F_0 \approx 0.004$meV.  Then identifying $(8F_0\langle |x_{\rm AF}|^2
\rangle)$ as the renormalization of $T_{\rm CO}$ we get
\begin{eqnarray}
\Delta (T) \approx \Delta  - 0.032 {\rm meV} \langle |x_{\rm AF}^2| \rangle 
\approx 0.04 {\rm K}  - (0.4 {\rm K}) \langle | x_{\rm AF}^2 | \rangle \ .
\end{eqnarray}
which is enough to shift ordering from F to AF at $T_{\rm CO}$,
where $\langle |x_{\rm AF}^2| \rangle \approx 1/2$. 

Finally, we should mention that this frozen phonon occurs whether or not the
CO phase is commensurate because its origin is in a coupling of the form
\begin{eqnarray}
V \sim u({\bf q}=0) \sigma({\bf q}) \sigma(- {\bf q}) \ ,
\end{eqnarray}
where $\sigma({\bf q})$ is the CO order parameter.

\section{The Magnetic Phase Transition}

\subsection{Phase of the R3 Structure}

We now discuss the magnetic  phase transition at which SO 
appears.  At first we neglect the fact that the system is
a mixture of spins of magnitude 2 and spins of magnitude 5/2 and we assume
that the uniaxial anisotropy aligns the spins along the ${\bf c}$ axis.
Then one introduces the local order parameter $S({\bf r})$ as the thermal
average of $S_z({\bf r})$, the $z$-component of spin at the site ${\bf r}$.
Also, as noted in Ref. \onlinecite{CHRIST}, if one neglects the coupling
between spin and charge, the symmetry of the SO free energy is the same
that of the CO free energy of Eq. (2).  Thus, if the transition is assumed
to be continuous, the ordering wave vector for this transition should be
unstable relative to the X point, just as we argued (in connection with
Fig. 6) in the case of the CO transition. In that case the representation
analysis of Ref. \onlinecite{CHRIST} for the wave vector $(1/3,1/3,0)$
is not definitive.  However, we temporarily overlook
the possible instability of the X wave vector and apply Landau theory to
the phase transition as if this wave vector were stable.  (In Appendix A
we discuss some difficulties in applying representation analysis to this
transition.) Therefore we write the free energy in terms of
$S_1({\bf q})=S_2({\bf q})\equiv S({\bf q})$, where the subscript labels
the two Fourier components of the unit cell.  We have that
\begin{eqnarray}
F = (1/2)(T-T_{SO}) |S({\bf q})|^2 + u |S({\bf q})|^4 + v |S({\bf q})|^6
+ \dots + w S({\bf q})^6 + w^*{S({\bf q})^*}^6 \ ,
\label{EQF1} \end{eqnarray}
where $T_{SO}$ is the magnetic (SO) transition temperature and we will set
\begin{eqnarray}
S({\bf q}) &=& | S({\bf q})| e^{i \phi} \ .
\end{eqnarray}
Under inversion symmetry $S({\bf q}) \rightarrow S({\bf q})^*$, which
implies that $w$ in Eq. (\ref{EQF1}) is real.
The last term in Eq. (\ref{EQF1}) is the lowest order one that fixes the
phase $\phi$ of the order parameter.  (It should be noted that it
is not easy to fix the this phase using only scattering data.)
There are two cases:[\onlinecite{RHO}]
\begin{eqnarray}
\phi = n \pi /3 \ \ \ {\rm if}\ w<0\  ; \hspace{1in} \phi= (n+1/2)\pi /3 \ 
\ \ {\rm if} \ \ w > 0\ ;
\label{PHASEEQ} \end{eqnarray}
with the results for the amplitudes in the magnetic unit cell
as given in the caption to Fig. 5. To determine the net moment of these
structures it is necessary to analyze the admixture of wave vector
$(0,0,0)$.[\onlinecite{CERR}]
Such an admixture comes about because ${\bf q}=(1/3,1/3,0)$,
is 1/3 of a reciprocal lattice vector and this fact allows an additional
term, $V$, in the free energy, where
\begin{eqnarray}
V &=& - S(0,0,0) \Bigl[ a S( 1/3,1/3,0)^3
+ a^*{S( 1/3,1/3,0)^*}^3 \Bigr] + \frac{1}{2} \chi^{-1} S(0,0,0)^2 \ ,
\end{eqnarray}
where $\chi$ is a stiffness (which is nearly temperature independent)
and $a$ is a constant which must be real in view of inversion symmetry.
Minimizing $V$ with respect to $S(0,0,0)$ we find that
\begin{eqnarray}
S(0,0,0) = a \chi [ S(1/3,1/3,0)^3 + {S(1/3,1/3,0)^*}^3] .
\end{eqnarray}
If $w>0$, then Eq. (\ref{PHASEEQ}) indicates that $S(0,0,0)$ is zero,
whereas for $w<0$ the system has a nonzero net moment, $M$.  The
early data of Ref. \onlinecite{IIDA} suggests that $M$ is nonzero.
However, recently we have learned[\onlinecite{PC1}] that the
system is more likely to have $M=0$, in which case we must
choose $w>0$. In this case one of the three sublattices is disordered.
(See the caption to Fig. \ref{FIG4L} with $\phi=\pi /6$.)
In this structure, all spins within a plane perpendicular to the
ordering wave vector have the same value, $a$, $-a$, or 0. 
This type of partial ordering was observed in the orientational ordering
of solid methane[\onlinecite{PRESS}], and, as in that case, we would not
expect a phase with partial disorder to continue to exist to arbitrarily
low temperature. In Appendix B we obtain $M(H)$ for this structure.

If, instead, the case $w<0$ is realized, then one would have
\begin{eqnarray}
S(0,0,0) = 2 a \chi S(1/3,1/3,0)^3 
\label{EQM} \end{eqnarray}
and if
\begin{eqnarray}
S(1/3,1/3,0) \equiv B \sim (T_{\rm SO} - T)^\beta
\label{EQN} \end{eqnarray}
then, within mean field theory
\begin{eqnarray}
S(0,0,0) \equiv A \sim (T_{\rm SO} - T)^{3\beta} \ ,
\end{eqnarray}
which gives an unusually large critical exponent for the magnetization.
For liquid crystals,[\onlinecite{AA}] this effect has been analyzed 
in detail within the renormalization group.
An effect similar to this has been seen for CO.[\onlinecite{COEXP}]

Next we discuss the magnetic eigenvector which was chosen in Ref.
\onlinecite{CHRIST}  to be $[1,1]$ (to best fit the experimental data).
With this choice of eigenvector
the spins form planes (perpendicular to the wave vector) of spins with
amplitudes proportional to $0$, $-1$, and $1$.  How is this choice
to be justified within Landau theory? In the `minimal' model used for CO,
one sees that for $\rho=0$ (the commensurate case) $F_{12}$=0
and one has isotropy in $S_{1z}$, $S_{2,z}$ space so that the eigenvector
can be $[\cos \theta ,\sin \theta]$ with any choice for $\theta$.  One way
to explain that the eigenvector is $[1,1]$ is to invoke an interaction which
tends to make the two spins in the rhombohedral unit cell parallel, so
that $F_{12}$ is negative real.  Since the spins are aligned along
${\bf z}$, the dipole interaction could accomplish this. However,
the energy $V$ of this interaction is probably too small:
\begin{eqnarray}
V &=& - \frac{2 g^2 \mu_B^2 S^2}{r^3} = - 0.045 {\rm K} \ ,
\end{eqnarray}
where we set $g=2$, $S=5/2$, and $r=2d'+d=14.1 \AA$ and combine with
a second term for which $r=d'+2d=11.1 \AA$.  Alternative
mechanisms to stabilize the antiferromagnetic spin structure involve
thermal fluctuations or (as we discuss below)
the distortion due to the frozen E$_g$ phonon.

Finally we discuss the diffraction at half integer values 
of $L$ which has a magnetic signature and which requires positing a magnetic
unit cell which is doubled along the ${\bf c}$ direction. As
stated in Ref. \onlinecite{CHRIST} one can think of this additional
diffraction as being due to ``the charge ordering, which decorates
the lattice with differing magnetic moment on the Fe$^{2+}$ and
Fe$^{3+}$ sites..."  
This effect can be seen within Landau theory as follows.  We introduce an
additional free energy $V$ of the following form, consistent with symmetry,
\begin{eqnarray}
V &=& - a \sum_{\bf r} x({\bf r}) S_z({\bf r})^2 \ ,
\label{VV1EQ} \end{eqnarray}
where $a$ is a constant.  The effect of this term is to increase (decrease)
$S_z({\bf r})^2$ when the site is occupied by an Fe$^{3+}$ (Fe$^{2+}$) ion.
In Fourier transform language this is
\begin{eqnarray}
V &=& - a \sum_{n; {\bf q} , {\bf k}_1 , {\bf k}_2} x_n({\bf q})
S_n({\bf k}_1) S_n ({\bf k}_2) \Delta({\bf q}, {\bf k}_1, {\bf k}_2) \ ,
\label{VVEQ} \end{eqnarray}
where $\Delta$ enforces wave vector conservation modulo a reciprocal lattice
vector and the subscripts label the Fe sublattices.
The term we focus on here involves charge ordering, so
that ${\bf q} = (1/3,1/3,1/2)+ \deltav$, where $\deltav$ is the
incommensurability. The magnetic variables then
can involve the wave vectors ${\bf k}_1 = (1/3,1/3,-1/2)-\deltav$ and
${\bf k}_2 = (1/3,1/3,0)$. Then we see that this interaction
couples $S(1/3,1/3,0)$ and $S(1/3,1/3,-1/2)^*+\deltav$.  Thus the
critical magnetic eigenvector is a mixture of these two variables.
This corresponds exactly to the idea of ``decoration," but it is hard
to estimate the importance of this effect.

\subsection{Removal of Frustration}

To develop further insight into this frustrated spin system, it is useful
to recall the results for the magnetic structure of the rhombohedral
$\beta$-phase of solid oxygen whose structure only differs from LFO
in that there are no bilayers: all TLL's are equally spaced.  (For a review
see Ref. \onlinecite{REVIEW}.)  A convincing theoretical analysis based
on quantum spin-wave theory was given in Refs. \onlinecite{ENRICO1} and 
\onlinecite{ENRICO2}.  However, when various theoretical results were
experimentally tested[\onlinecite{DUNST}], it was not entirely clear
which theoretical model was most appropriate for $\beta$-oxygen. In any
event, the magnetic correlation length is so short ($5\AA$)[\onlinecite{DUNST}]
that is seems unrealistic to speak of any long range order.

\begin{figure}[h!]
\begin{center}
\vspace {0.4 in}
\includegraphics[width=3.0 cm]{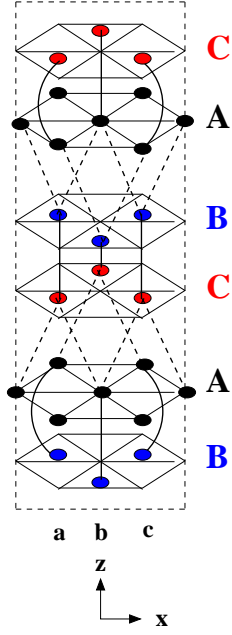}
\caption{\label{INTFIG} (Color online)
Additional exchange interactions attributed to the transverse frozen phonon.
Dashed (full) lines denote  additional antiferromagnetic (ferromagnetic)
interactions. The values of $S_z$ for all sites in the same $y$-$z$ plane
are $a$, $b$, and $c$ as indicated, where $b=-a$ and $c=0$, in the
notation of Eqs. (\ref{EQM}) and (\ref{EQN}). It is assumed that the in-plane
interactions are dominant.}
\end{center}
\end{figure}

Accordingly, a central open question is to explain why the SO
in LFO is so different from that of $\beta$-oxygen.  One possibility
is that the small distortion which we invoked to explain the
crossover from FI to AFI might introduce small addition exchange
interactions which remove the frustration of the rhombohedral
antiferromagnet. To explore this possibility we write
\begin{eqnarray}
J_n &=& J_n^{(0)} + \delta J_n \ ,
\end{eqnarray}
where the $J_n^{(0)}$'s are the exchange interactions which are
consistent with the R$\overline 3$m symmetry.  In Fig. \ref{INTFIG} we show
a set of interactions $\delta J_n$ which have the correct symmetry to
be induced by the frozen E$_{\rm g}$ phonon and which, if they are dominant,
resolve the frustration. Note that even though CO is incommensurate, the
frozen phonon is commensurate.  So here we are considering a commensurate
effect of incommensurate charge ordering.

To analyze this possibility, we replace the $J_3$ interaction by 
$J_3(1+ \epsilon)$ for the dashed bonds in Fig. \ref{INTFIG}
and the other $J_3$ interactions by $J_3(1-\epsilon)$.  Similarly, we
replace the $J_2$ interaction by $J_2(1-\epsilon)$ for the solid bonds in
Fig. \ref{INTFIG} and the other $J_2$ interactions by $J_2(1+\epsilon)$.
Then Eq. (\ref{EQF12}) remains valid but now
\begin{eqnarray}
\Lambda &=& (1+\epsilon) e^{iaq_y\sqrt 3 /6} \cos(aq_x/2)
+ (1-\epsilon) e^{-iaq_y\sqrt 3 /3} \nonumber \\
&=& -2 \epsilon - (\sqrt 3 /2)(\rho_x+i\rho_y) - \epsilon \rho_x \sqrt 3 /2
+ i \epsilon \rho_y \sqrt 3 /6 + {\cal O}(\rho^2)\ .
\end{eqnarray}
To maxmimize $|F_{21}|$ (for $U_2U_3>0$)
set $\exp(icq_z/3)= \Lambda/\Lambda^*$, so that
\begin{eqnarray}
|F_{21}| &=&  (U_2+U_3) |\Lambda| \ .
\end{eqnarray}
Then the minimal eigenvalue is found by minimizing
\begin{eqnarray}
\mu (\rho_x, \rho_y) &=&
T-3U_1 + (3/4)U_1 \rho^2 \nonumber \\ && - (U_2+U_3) \Biggl(
[ 2 \epsilon + (1+\epsilon)(\sqrt 3/2)\rho_x]^2 + (3/4) (1 - \epsilon/3)^2 
\rho_y^2 \Biggr)^{1/2} \ .
\end{eqnarray}
When $\epsilon=0$, this is a function of $\rho_x^2 + \rho_y^2$ and
is consistent with the existence of a degeneration line.  However,
when $\epsilon$ is nonzero, then the minimum occurs for $\rho_y=0$
(so that $q_z=0$) and
\begin{eqnarray}
\rho_x &=& (1+\epsilon)(U_2+U_3)/(\sqrt 3 U_1) \ .
\end{eqnarray}
One might wonder where the extrema we found for CO at $cq_z= 2p \pi$, for
$p=1$ and $p=2$ have gone. The answer is that there are three CO domains
corresponding to which there are three distortions, the transverse
phonon displacement being perpendicular to the in-plane projection of
the ordering wave vector.  So we have three different SO domains, each one
tied to one of the three possible CO domains.

An important question is: since the wave vector is not at a special,
high-symmetry point (see Fig. 6), the wave vector should not be
commensurate if the SO transition is a continuous one.  It is not clear
that the sensitivity of the neutron diffraction experiment of Ref.
\onlinecite{CHRIST} is sufficient to detect the very small
incommensurability that might attend this magnetic transition.
It would be of interest to have a high precision determination of
the SO wave vector, to check whether it is or is not commensurate.

If the SO phase is truly commensurate, then one would have to entertain a
scenario to accommodate such a fact.  The one scenario that is excluded
is that the commensurate SO state is reached via a single continuous phase
transition.  Possibly there are two phase transitions, the first, in which 
there develops incommensurate order, followed by a second one 
into the commensurate antiferromagnetic state.[\onlinecite{DS}]
The presence of two nearby transitions in parameter space
would seem to signal a nearby multicritical point at which the
two transitions coincide.  To check for that, it would be useful
to have very precise measurements of the specific heat and
susceptibility to get the critical exponents that characterize this 
transition.  A different scenario is that the magnetic transition
is a first order one to a commensurate SO phase.

\subsection{Field Cooling}

Finally, we mention the intriguing data of Fig. 3c of Ref. \onlinecite{WEN},
where it is shown that cooling in a field along $(1,-1,0)$
from $T>T_{CO}$ causes a
pronounced reduction in the CO diffraction at 300K.  This data
raises a natural unanswered question: does this `missing" intensity
in the AFI scattering show up as new FI scattering at
$(1/3,1/3,n)$ for integer $n$.  If so, it would mean that the
magnetic field could stabilize the FI state for $T<T_{CO}$ and
it would be of interest to know whether such a state was or was not
commensurate.

\begin{figure}[h!]
\begin{center}
\vspace {0.4 in}
\includegraphics[width=5.0 cm]{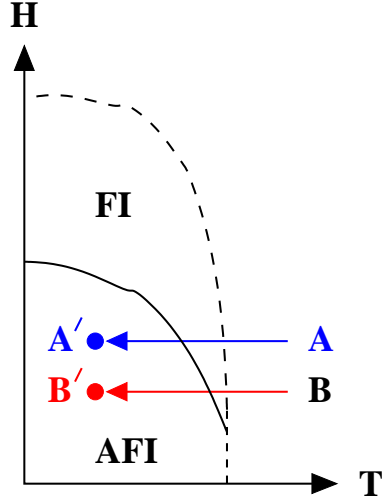}
\caption{\label{PHASE} (Color online)
Proposed phase diagram for CO as a function of maganetic field $H$
and temperature $T$.  Dashed lines represent continuous transitions
and the solid line a first order transition.}
\end{center}
\end{figure}

Here we present a partial explanation for the above field cooling
scenario.  We start by noting that in Ref. \onlinecite{WEN} it
is shown that application of a magnetic field in the plane of the
TLL's decreases the AFI correlation length.  This suggests that
such a magnetic field tends to destabilize the AFI phase, possibly 
making the FI phase relatively more stable.  If this is the case,
the one might have a phase diagram like that shown in Fig.
\ref{PHASE}. Then in the various scenarios of cooling one would
start in the disordered phase at points like A and B and cool to points
A' and B'.  Clearly, if this is done at zero field and then a field
not large enough to go into the FI phase is applied, no dramatic field
dependence will be observed, in agreement with their observations.
However, if one starts from point like
A or B, then when the system is cooled it passes though a region of
FI ordering which then can be supercooled while reaching the final
points A' or B'.  Then, as a function of time
the system would evolve in some irregular
process into the equilibrium state of AFI order.  This might happen
without displaying a dramatic dependence on the field-cooled
value of the magnetic field. Indeed the data shows that after
a sharp decrease for very small field, the resulting AFI order
does not depend strongly on $H$.  This proposal suggests that
it would be useful to monitor the $(1/3,1/3,0)$ reflection
under field cooled conditions (to see if the decrease in intensity
at $(1/3,1/3,3/2)$ is accompanied by an increase at $(1/3,1/3,0)$.
It would indeed be interesting if an in-plane magnetic field could
stabilize a nonzero polarization. Note that an in-plane magnetic field
may be more effective than one parallel to the $c$-axis, because the
perpendicular susceptibility is usually larger than the parallel 
susceptibility.

\section{SUMMARY}

Here we briefly summarize our conclusions.

$\bullet$
We show that the appearance of an incommensurate wave vector for
charge ordering is a result of symmetry (or more accurately,
due to a lack of symmetry).

$\bullet$
By comparing our theory with experiment we have assigned values to
several of the phenomenological charge-charge interactions. In particular,
the signs and magnitude of the incommensuration and the fact that
ferroelectric fluctuations dominate in the disordered phase are
explained by simple choice of these interactions.

$\bullet$
The cross over, as the temperature is lowered through the
charge ordering temperature, from ferroelectric to antiferroelectric
incommensurate structure can be explained by the temperature
dependent renormalization of the transition temperature
due to charge-phonon coupling.  

$\bullet$
We have performed a first-principles calculation of the zone
center phonon energies (assuming no ordering of charge or spin).
The phonon with the correct symmetry to couple effectively to
the charge ordering has a rather low (20 meV) energy, corresponding
to the sliding (transverse to the ${\bf c}$ axis) of alternate
Fe layers with respect to one another.

$\bullet$
We have developed a Landau theory which describes the phase of
the recently observed spin ordered state having zero net magnetic moment.

$\bullet$
In principle, if the spin ordering transition is continuous, the
spin ordered phase would be expected to be incommensurate.  This
suggests the need for a high precision determination of the spin
ordering wave vector to check whether it is or is not commensurate.
If the spin ordered phase truly is commensurate, then it would be of interest
to investigate the scenario of ordering, which can not be via a single
continuous transition.  One way to pin down the scenario would 
be to determine the critical indices $\alpha$, $\beta$, and $\gamma$,
associated respectively with the specific heat, the magnetic order
parameter, and the susceptibility.  

$\bullet$
We have also suggested experiments to test our proposal that
the sharp decrease in intensity of antiferroelectric
charge scattering as a function of magnetic field in the field cooled
scenario might indicate that the magnetic field tends to stabilize
ferroelectric charge ordering and possibly a consequent polarization.

\vspace{0.2 in} \noindent
ACKNOWLEDGEMENTS
We thank A. D.  Christianson and M. Angst for helpful correspondence,
and A. Boothroyd for introducing us to this subject.  We also thank
Q. Xu, S. Shapiro, D. Singh, E. J. Mele and T. C. Lubensky for
stimulating discussions. We also thank E. Rastelli for a discussion
of Refs. \onlinecite{ENRICO1} and \onlinecite{ENRICO2}.

\begin{appendix}
\section{Representation Theory}
In Refs. \onlinecite{CHRIST} and \onlinecite{ANGST} representation theory
is used to analyze possible magnetic ordering patterns and charge ordering 
patterns, respectively.  In their approach, they implicitly assume that
the wave vector at the appropriate X point is stable with respect to
the addition of further neighbor interactions.  As we have seen in Sec. 
II, this assumption is not actually valid, especially for CO.
To see this explicitly, consider the structure of the two by two matrix
$F_{nm}$ of Eq. (4) which determines the eigenvectors.  Exactly at the
X point and when arbitrary interactions are allowed, $F_{21}$ is scaled
by the interactions $V_{1,2}$ between sites \#1 and \#2 which are displaced
from one another by a vector along the $c$-axis.  This interaction is
extremely small, since it connects sites which are not in adjacent
bilayers, but are in second (or further) neighboring bilayers.
Representation theory bases the eigenvector equation on this symmetry 
and leads to eigenvectors that are either even or odd under inversion.

However, as noted in Sec. II, this type of analysis is invalidated by the
fact that for LFO the X point is not actually stable. For wave vectors
near the X point, we explicitly displayed in Eq. (7) the term in
$F_{12}$ which is linear in the displacement from the X point.
The interactions $U_2$ and $U_3$ which scale this linear term are 
very much larger than $V_{1,2}$ whose existence is ignored if representation
theory is invoked for the commensurate case.  The major effect of
this linear term is that the eigenvectors, instead of being even and odd,
as in the analysis of Refs. \onlinecite{CHRIST} and \onlinecite{ANGST},
are now complex and are determined by the phase of $F_{21}$ given 
by Eqs. (4) and (7).  Note that this phase will, in general, be
different for each of the three domains, and inclusion of the correct
phases, might affect the determination of the domain populations.

\section{Equation of State for the Antiferromagnet}

\begin{figure}[h!]
\begin{center}
\vspace {0.4 in}
\includegraphics[width=5.0 cm]{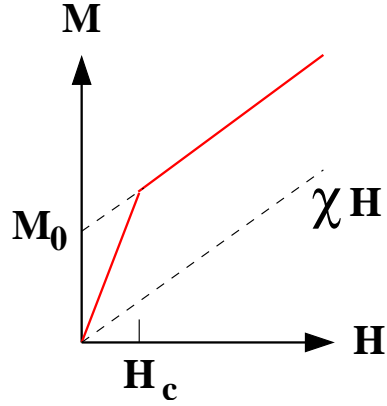}
\caption{\label{CHIH} (Color online)
The $c$-component of magnetization $M$ (solid line)
versus the magnetic field $H$ along
${\bf c}$ for the antiferromagnetic phase. Here $M_0=2a\chi \sigma^3$
and $H_c$ is given by Eq. (\ref{HCEQ}).  Near $T_{\rm SO}$ both $M_0$
and $H_c$ are of order $(T_{\rm SO}-T)^{3/2}$.}
\end{center}
\end{figure}

In this appendix we obtain $M(H)$, where $M$ is the net magnetization
(along the ${\bf c}$ axis) for the model of Eq. (\ref{EQF1}) and $H$ 
is the external field applied parallel to the $c$ axis. Accordingly we
add to the free energy the term $- HM$, where $\chi$ is the parallel
susceptibility, so that with $|S({\bf q})| \equiv \sigma$, we have
\begin{eqnarray}
F = \frac{1}{2} (T-T_{SO}) \sigma^2 + u \sigma^4 
+ 4w \sigma^6 \cos^2(3 \phi) -2 a \sigma^3 M \cos (3 \phi)
+ \frac{1}{2} \chi^{-1}M^2 - MH \ ,
\end{eqnarray}
where $\chi$ is the parallel susceptiblity and we kept $\phi$-independent
terms only up to order $\sigma^4$ because our analysis is not valid when
$T \ll T_{SO}$.  Minimizing with respect to $M$ yields
\begin{eqnarray}
M &=& \chi [ H + 2a \sigma^3 \cos (3 \phi)]
\end{eqnarray}
so that
\begin{eqnarray}
F &=& \frac{1}{2} (T-T_{\rm SO}) \sigma^2 + u \sigma^4 + 4 w \sigma^6 \cos^2
(3 \phi) - \frac{1}{2} \chi [ H + 2 a \sigma^3 \cos(3 \phi)]^2 \ .
\end{eqnarray}
When this is minimized with respect to $\phi$ we find two regimes:
\begin{eqnarray}
H>H_c \ : \hspace{1 in} M &=& \chi \Bigl[ H + 2a \sigma^3 \Bigr] \nonumber \\
H<H_c \ : \hspace{1 in} M &=& \chi H \Bigl[ 1 + \frac{2a^2 \chi}
{4w-2a^2 \chi} \Bigr] \ ,
\end{eqnarray}
where
\begin{eqnarray}
H_c &=& \frac{(4w-2a^2 \chi) \sigma^3 } {a \chi} \ .
\label{HCEQ} \end{eqnarray}
which leads to the $M$ versus $H$ curve shown in Fig. \ref{CHIH}.
The value of $\sigma$ is approximately $[(T_{\rm SO}-T)/(4u)]^{1/2}$
for $T$ near $T_{\rm SO}$.

\end{appendix}


\begin{thebibliography} {99}
\bibitem{REY} %17
Y and Ikeda[\onlinecite{IKEDA2}]
also observe diffraction at half-integer values of $\tilde q_z$, but
seem not to realize that this corresponds to a doubling of the unit cell.
\bibitem{LAMBDA} %23
In the presence of an interactions between sites \#1 and \#2 within the
same unit cell, the form of $\Lambda$ is $\Lambda = \Lambda_0 + A \rho$,
where $\Lambda_0$ and $A$ are constants.  The presence of a term linear
in $\rho$ guarantees that the X line is unstable.
\bibitem{rho2} % 24
The values for the $U$'s are deceptively small because when $\sigma$'s
are equal to 1, they represent the interaction between two effective
charges $e/2$, the difference between the local charge and the
average valence charge.
\bibitem{rho} % 25
Our $\rho$ is related to Y's $\delta$ by $\delta = 4 \pi \rho /3$.
Y gives $\delta = 0.0081$.
\bibitem{EXPL} %27
C. Broholm and D. Reich have pointed out to us that the $U$'s will show
a temperature
dependence because of the temperature dependence of the  dielectric
constant.  That is certainly the case near or below 200 and may be
quite relevant for the nature of the transition near 170K.  However,
the amplitude of the zig-zag pattern of diffraction guarantees that
screening is very strong near and above the CO transition and there
is no evidence that the temperature dependence of the dielectric
constant causes the cross-over from FI to AFI fluctuations near
$T_{\rm CO}$ where the dielectric constant is large.
\bibitem{FN3} %32
If we were to invoke a perturbation {\it linear} in $x$
(which would result from a Lu-Fe interaction), then we would obtain
Eq. (\ref{EQFF}) without the factor $\langle x({\bf q}) x({\bf q})^* \rangle$,
and we would have to introduce an ad hoc temperature dependence to $\Delta$.
\bibitem{CERR}
Contrary to the statement in Ref. \onlinecite{CHRIST} that the above
analysis (assuming $w<0$) ``results in a ferrimagnetic structure as
shown in Fig. 3...", one can only get a nonzero net moment from a
nonzero ordering at zero wave vector, as is done here.
\bibitem{COEXP} %38
In Fig. 4d of Ref. \onlinecite{ANGST} one sees the intensity of
a ninth overtone, whose critical exponent within mean field
theory would be $9 \beta$.  The data is obviously not sensitive
enough to test this prediction of the exponent.
\end{thebibliography}
\end{document}